\documentclass[vecphys]{svmult}

\usepackage{makeidx}         % allows index generation
\usepackage{graphicx}        % standard LaTeX graphics tool
\usepackage{multicol}        % used for the two-column index
\usepackage[bottom]{footmisc}% places footnotes at page bottom
\makeindex             % used for the subject index
\begin{document}

\title*{Geometric Algorithms for Identifying and Reconstructing Galaxy Systems}
\titlerunning{Geometric Cluster-Finding Algorithms} 
\author{Christian Marinoni}
\institute{Centre de Physique Th\'eorique, University of Provence, UMR 6207\\ 
Campus de Luminy, 13256 Marseille, France \\
\texttt{marinoni@cpt.univ-mrs.fr}}
\maketitle

\section{Introduction}
\label{sec:1}

The recognition of patterns and structures in a given point distribution is 
common challenging task in science and engineering.
In astronomy, in particular, 
the night-sky itself offers a natural  set of points suitable 
for topological analysis.
Ancient astronomers grouped bright stars into simple patterns (constellations)
whose form and configuration could be easily identified. 
For a long time, this  simple eyeball classification  
has provided us with useful signposts for 
tracking the flow of the seasons or for orienting sea travellers.

Modern astronomers detect and catalogue the 
structures traced by galaxies on the grand scale of the universe, as well. This activity 
has actually become a discipline on itself called cosmography.
However, since the human eye and human mind respond in a biased way to contrast and
continuity, cosmography is not based on visual impressions, as in the old days,  
but on a quantitative statistical description of patterns.
What has remained unchanged across time is the importance of this activity.
For example, by measuring in an objective and reproducible  way  the large-scale 
spatial arrangement of galaxies we can have access  to fundamental  information about 
the universe's mass content and distribution. Additionally,  
an unbiased reconstruction of the galaxy distribution
provides us with a quantitative characterization of the environment in which 
galaxies live, i.e. groups, cluster, super-clusters, filaments and walls. 

Groups and clusters of galaxies, in particular, provide ideal laboratories for studying 
many aspects of the physics of galaxies within a well-defined, controlled, 
environment. Therefore, the finer their identification and reconstruction, 
the finer the scientific issues one can resolve. 
For example, one can  evaluate the effects of the 
environment on the evolution of galaxies and assess which physical mechanisms  
(for example ram pressure stripping of gas, galaxy-galaxy merging, etc.) 
are (or not) crucial in determining the present  day aspect 
of galaxies. Answering these key topics will provide us with insights concerning
the nature of galaxy evolution itself and will clarify 
whether galaxy properties  were established early on when galaxies first assembled
(the so-called ``nature" hypothesis), 
or whether they are the present day cumulative end product of multiple environmental processes
operating over the entire history of the universe (the ``nurture" scenario).

The specific  theme of this review is to describe  various algorithms developed 
by cosmologists for identifying and reconstructing groups and clusters of galaxies 
out of the general galaxy distribution.  To this aim, I will  follow the progression of
 clusters detection techniques 
through time, from the very first visual-like algorithms to the most performant 
geometrical methods available today.
This will  allow readers to understand the development of the field as well as the 
various issues and pitfalls  we are confronted with. 
In particular, I will emphasize  some  recently developed, optimal 
detection techniques which are based on the Voronoi and Delaunay geometric models. I will 
overview their relative strengths
and limitations and compare their performances with the  more standard 
cluster-finding tools.

This paper is structured as follows: 
in \S 2 I will introduce the notion of galaxy groups and clusters,
briefly presenting their main astrophysical properties. 
In \S3 I will survey some  general group/cluster identification tools 
traditionally used by astronomers for detection  in both 2 and 3 dimensional 
space. Voronoi-Delaunay based cluster-finding algorithms are presented in \S 4.
Conclusions are drawn in \S 5.

\begin{figure}
\centering
\includegraphics[width=12cm]{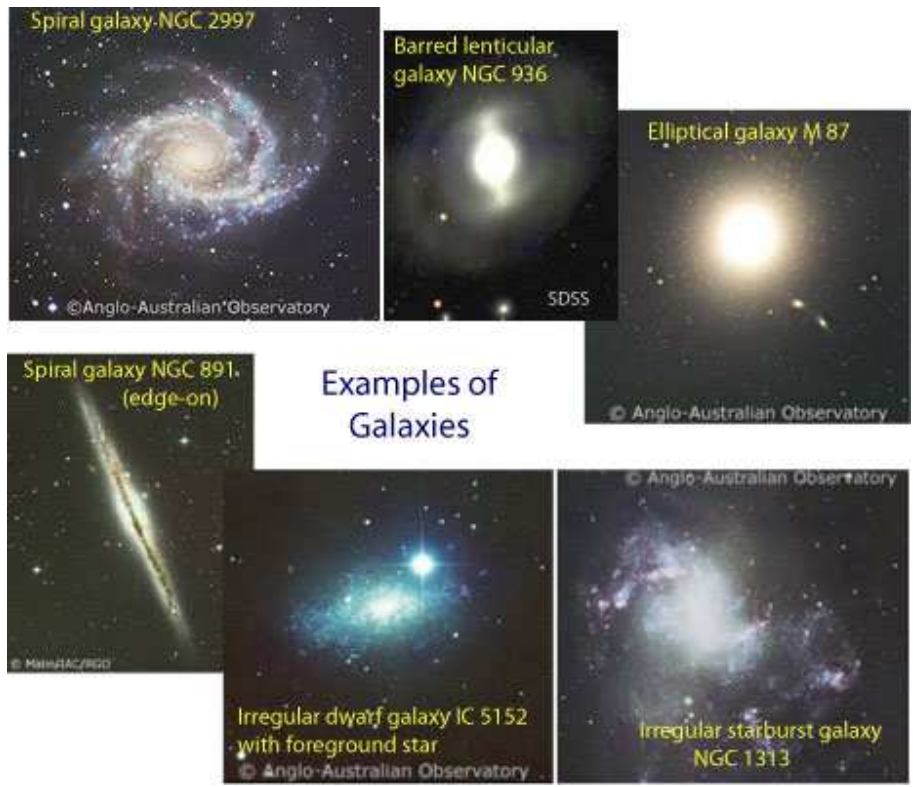}
\caption{Galaxies have a variety of morphologies, stellar content and kinematics. 
Elliptical galaxies are pressure-supported systems which   
contain an old population of stars having red colors (cold 
surface temperatures). The fundamental building blocks 
of centrifugally supported spirals and dynamically irregular galaxies  are younger 
and hotter (bluer) stars.
(Credit: Anglo-Australian Observatory, D. Malin, IAC, RGO and SDSS)}
\label{fig:types}       %  
\end{figure}

\begin{figure}
\centering
\includegraphics[width=12cm]{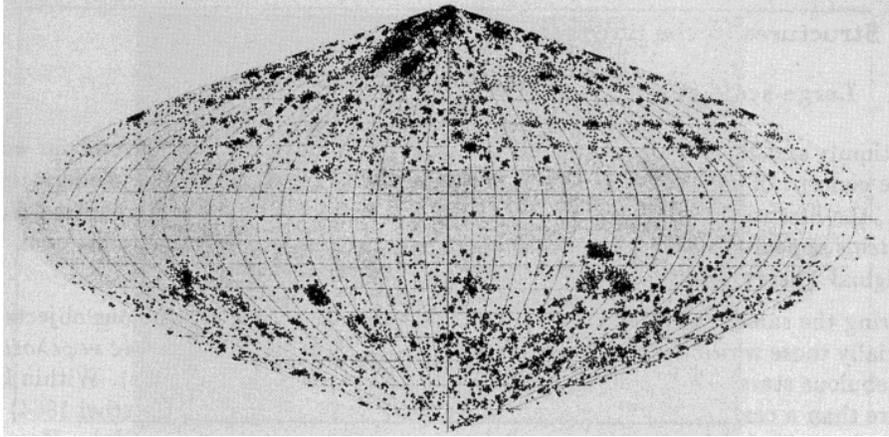}
\caption{Charlier's  map of  11,475 nebulae \cite{charly},
based on the  New General Catalogue and  the two Index
Catalogues (\cite{D1},\cite{D2},\cite{D3})}
\label{fig:charlier}       %  
\end{figure}

\begin{figure}
\centering
\includegraphics[height=10cm]{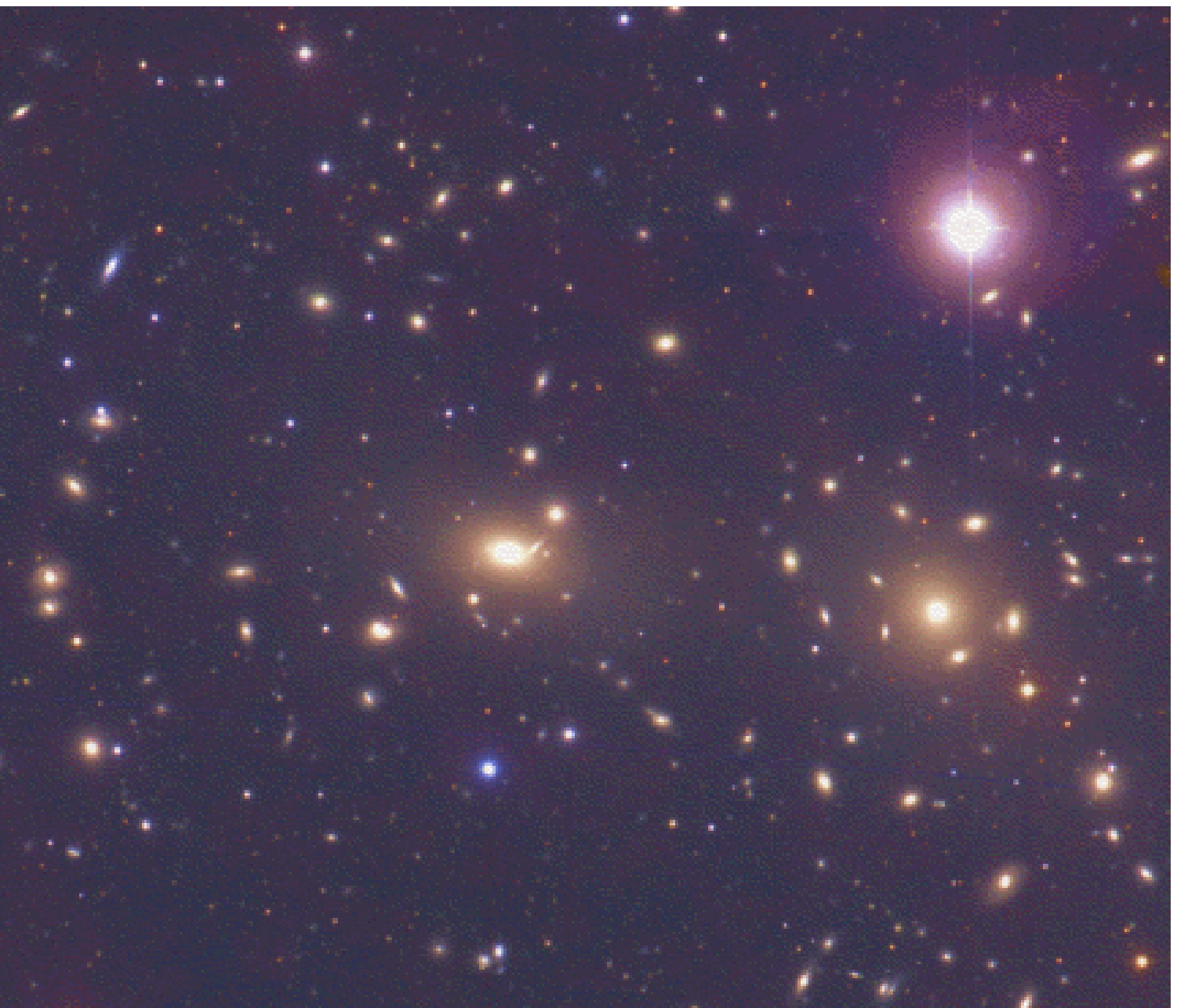}
\caption{The Coma cluster - Abell 1656 - is an example of a rich regular cluster. 
Here the center of the cluster is shared between the two bright elliptical galaxies
 (the two bright objects at the center of the image).
This cluster is 7Mpc in diameter and is thought to contain almost 2000 galaxies. The 
bright object in the upper right is a star from our own galaxy.
Despite its regularity Coma is actually a merger of at least three smaller groups
(e.g. \cite{ada}).
Irregular clusters can also have a prominent bright central galaxy  (and can sometimes 
be a giant elliptical galaxy) but are more disorganized in appearance. There is also no
 centralized concentration of galaxies near the center of the cluster.}
\label{fig:coma}       %  
\end{figure}
\section{What is a cluster?}

According to the relativistic theory of gravity, 
matter is smoothly distributed, anchored to space and expanding 
with the metric of the universe. In particular, on small cosmological 
scales, theory predicts \cite{desitter}  and observations confirms \cite{hubble} 
that the redshift \footnote{The  redshift between two objects 
(commonly called the {\it emitter} and the {\it observer})
is an astronomical observable  defined as the  relative shift of 
electromagnetic wavelengths due to the expansion of the space 
between the source and the observer 
\[z=\frac{\lambda_{o}-\lambda_{e}}{\lambda_{e}}\]} $z$
between  
any two given matter particles   is proportional to their
separation $r$ via the relation 

\begin{equation} 
cz=Hr
\label{eq1}
\end{equation}

\noindent where c is the speed of light and $H$ the Hubble constant.
The Doppler formula ($V=cz$) allows us to re-interpret the redshift 
as a measure of the 
recession velocity of galaxies in the local universe. As a result, the Hubble relation 
of eq. (1) which characterizes
the local expansion properties of the universe, also describes the apparent outward 
radial flow of galaxies as measured by a terrestrial observer.

Galaxies, however, the basic building blocks 
of the universe (see Fig. \ref{fig:types}), 
are not evenly distributed throughout the space. 
The variance of the counts in arbitrary cells randomly thrown in the universe is 
larger than what we would  expect from a purely Poissonian process. This is graphically seen 
in figure \ref{fig:charlier} which shows a map of galaxies derived by Charlier
(1922) \cite{charly}. Historically, this map represented
one of the first ever pictures of an all-sky distribution of galaxies 
(or nebulae as they where called at 
that time). Commenting this plot, Charlier wrote: {\em a  glance  at   this  plate  suffices  for
stating how the Milky Way, which is designed
by the great axis of the chart is systematically  avoided  by  the  nebulae. A
remarkable property of the image is that the
nebulae seem  to be  piled up  in clouds (as
also the  stars  in the  Milky Way).  Such a
clouding  of  the  nebulae  may  be  a  real
phenomenon, but it may also be an accidental
effect...}.

We now know that such a ``clouding of nebulae" is not a spurious projection
effect. Local gravitational perturbations tend to make receding galaxies slow down and clump 
together in small groups and sometimes in enormous complexes. Major collections
(up to several thousands) of galaxies are called galaxy clusters. Actually, the
universe is not composed of two  distinct classes of objects:
single galaxies and galaxies in groups.
The long range action 
of the gravitational field introduces a strong correlation in the matter density
field on scales less than $5$ Mpc.  As a result, and as far as the local universe is concerned, 
galaxies are preferentially found in structures ranging from pairs
(or binary systems) and triplets, through small groups,  up to rich (and rare) clusters.
Moreover, clusters themselves are often 
associated within larger gravitational structures called  super-clusters.
In this picture virtually no galaxies in the universe can be considered to be truly isolated.

\begin{figure}
\centering
\includegraphics[height=8cm]{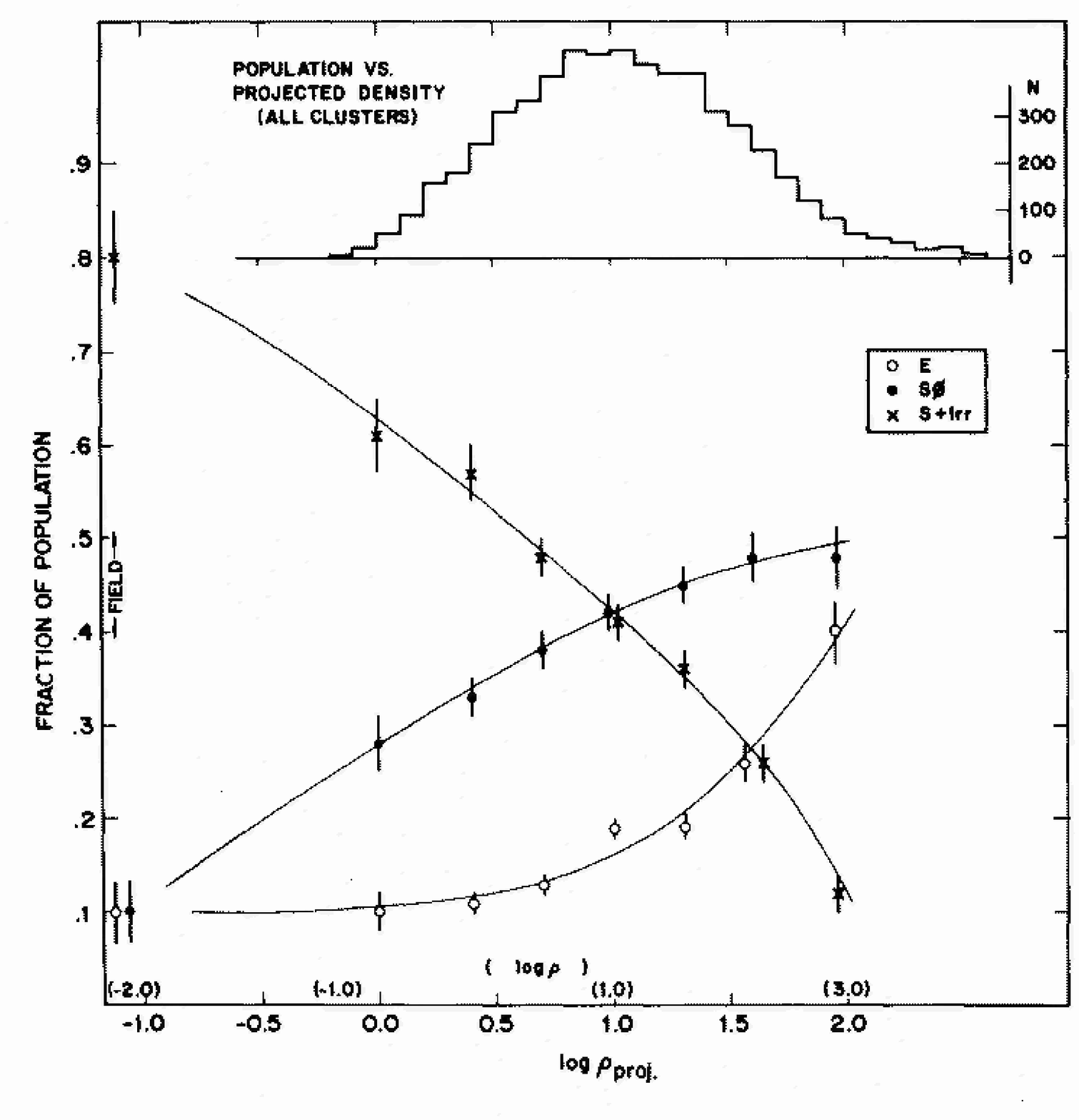}
\caption{The morphology-density relation  
The fraction of ellipticals (old star galaxies), lenticulars and  spirals 
(young star  galaxies) as a function of the projected density.
The upper histogram shows the number distribution of 
the galaxies over the bins of projected density.
(Taken from \cite{dressler}.)}
\label{fig:md}       %  
\end{figure}

In models for the gravitational formation of structure, 
the smallest density perturbations in an otherwise smooth universe collapse first 
and eventually build the largest structures, rich groups and clusters of galaxies. 
The most massive conglomerations of galaxies 
are therefore not only the largest 
gravitationally bound systems we know,  
but also  the most recent objects to have arisen in 
the hierarchical structure formation of the universe. 

A regular cluster can be described, at first order, as a spherically symmetric object 
in hydrostatic equilibrium whose members share the same average kinetic energy
(temperature). This simple model is able to capture most of its physical 
properties, in particular 
the smooth increase of the space density of galaxies towards the center of the structure.

Not all the groups of galaxies are isothermal sphere of gas.
A significant fraction of small groups are probably 
not bound nor relaxed 
structures but occur just by  chance alignments of galaxies; they form temporarily
but then dissolve as galaxies move past one another. 
As simulations offer more dynamical information than observations, one
can use N-body data to calculate whether the groups 
of galaxies are gravitationally bound objects. Niemi et al. \cite{sami}, for example, 
showed that about 20 per cent of nearby groups of galaxies, identified by the 
same algorithm as in the case of observations, are not bound, 
but merely groups in a visual sense.

Large clusters, on the contrary,  have radii up to 
2-3 Megaparsec, they contain from 50 to 5000 galaxies which move 
in the cluster deep gravitational potential  with projected
1D peculiar velocities (i.e. deviations from the smooth Hubble recession
flow of eq. \ref{eq1})  ranging from 400 km/s to 1500km/s
(see the review of \cite{bahcall}.)

Many independent evidences suggest that rich clusters have relaxed to 
a bound equilibrium configuration. This inference is confirmed for example by comparing 
the {\it crossing time} of a typical galaxy in the cluster with the age of the universe. 
The crossing time is defined by $t_{cr}=R/\langle v \rangle$ where R is the size of a cluster and $\langle v \rangle$ 
is the typical 1D peculiar velocity of its members. For the Coma cluster (see Fig. 3),
taking $\langle v \rangle=10^3$ km/s and $R=1$ Mpc, the crossing time is about one tenth the 
age of the universe. This is compelling evidence that the cluster must be a bound
system or else the galaxies would have dispersed long ago.

More than 70 years ago, however,  Zwicky realized that projected peculiar velocities 
as high as those observed ($\langle v \rangle \sim 10^3 km/s$) are far too large for clusters 
to remain gravitationally bound by the  mutual gravitational attraction of its visible 
galaxy members only.
As a matter of fact, most structures turn out to be unstable if the virial theorem 
is applied to their visible members. Let's discuss this paradox in more detail.
By count of the galaxies, one can determine the integrated
luminosity of a cluster and by applying the virial theorem to its visible members
one can also infer its mass. 
The inferred ration $M/L$ is about one order of magnitude larger than the
corresponding value for elliptical galaxies, the galaxy type with the largest 
$M/L$ (see Fig. \ref{fig:types}). 
Thus accounting only for the visible material contained in galaxies, clusters
turn out to be unstable. If, on the contrary,  clusters are relaxed structures, 
this result apparently implies that they must contain 
considerably more mass than is visible in galaxies - the dawn of the 
missing mass problem (see \cite{biviano} for an historical account).
This issue has by now been firmly established. We 
interpret these evidences assuming that galaxies are surrounded by huge
halos of exotic (and dark) form of matter. Stars in galaxies and hot diffuse intra-cluster gas
contribute less than about $20\% $  to the total mass of clusters.

Not only clusters contain exotic forms of matter, but also their galactic content 
is peculiar. The
overall mixture of galaxy types in clusters differs from that 
of the general field (see Fig. \ref{fig:types}) : whereas about 70$\%$ of the field (isolated) galaxies are spirals, 
clusters are dominated by ellipticals, whose relative abundance increases as a function 
of the mass of the system. In particular, 
the inner part (the core) of a typical rich cluster  
consists mainly of the brightest and most massive 
galaxies of the whole system. They are essentially 
ellipsoidal objects with an old (early-type) population  
of cold stars which mostly 
emits in the red part of the electromagnetic spectrum. 
These giants (called cD galaxies, see Fig. \ref{fig:coma}) have in general multiple nuclei and 
their stars are 
characterized by random, non coherent motions. This suggests that 
they might be formed as a consequence of fusion and merging of smaller galaxies 
in the core of the cluster.
On the contrary, in the external, low density regions of a cluster
the galaxy population is mostly dominated by bluer, gas rich, younger (late-type)  
galaxies such as spirals or irregulars (see Fig. \ref{fig:md}).

\begin{figure}
\centering
\includegraphics[width=8cm,scale=3,angle=-90 ]{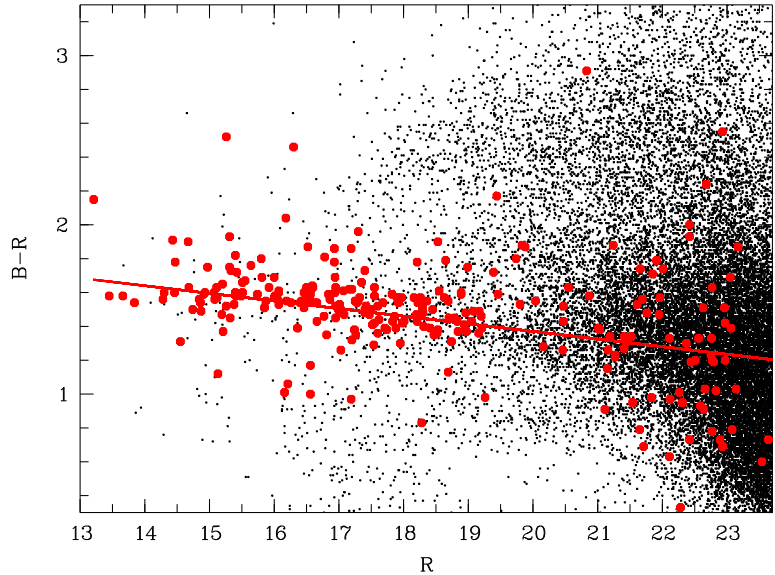}
\caption{The color-magnitude diagram of galaxies. Spectroscopically 
confirmed members of the cluster (Coma in this specific case) are indicated with red circles,
while black dots represent galaxies along the line-of-sight.
Clusters contain unusual concentrations of bright galaxies which look more redder than most field galaxies. These galaxies exhibit a tight correlation between their colors and magnitudes.
The narrow horizontal line of galaxies at nearly constant color is referred to as the elliptical ridge. Astronomers look for this characteristic shape in the color-magnitude  
diagram to select cluster members against foreground and background galaxies. 
(Taken from \cite{adami2}.)}
\label{fig:cm}       %  
\end{figure}

A key observational discovery
 was that all early-ype galaxies in a cluster have the same color, 
only weakly dependent on their 
luminosity (see Fig. \ref{fig:cm}). These  galaxies define a characteristic
nearly horizontal sequence in color-magnitude diagrams
\footnote{For historical reasons, and also to take into account
the specific eye response to illumination, optical astronomers evaluate 
the luminosity of an object using the notion of {\it magnitude} $m_x$. The 
photon flux $S_x$  collected in a given electromagnetic band $x$ is rescaled according 
to the relation
\[m_x=A_x-2.5 \log S\]
where  $A_x$ is a constant which depends on the specific filter in which light is collected.
This means that the larger the apparent magnitude of an object, the fainter the source is.
The difference between the magnitudes of the same object measured in different bands 
is called color of the object and it is a measure of the ratio of the fluxes emitted 
in different portion of the electromagnetic spectrum. A red color means that the flux 
emitted at longer optical wavelengths is larger than at smaller wavelengths.}
which is called the color-magnitude relation \cite{baum}  or red sequence (\cite{YGL}).
Moreover, in clusters with the same
redshift and mass, all early-type members have also the same colors. Comparing 
the red sequences of clusters at different redshifts, one finds that 
early-types are redder the higher the redshift is. This well defined red-sequence is 
of crucial importance for our understanding of the evolution of galaxies in clusters. It tells us
that the stellar populations in clusters have very similar ages. In fact the colors
of cluster members is compatible with their stellar population being roughly
the same age as the universe at that particular redshift.

\begin{figure}
\centering
\includegraphics[width=12cm]{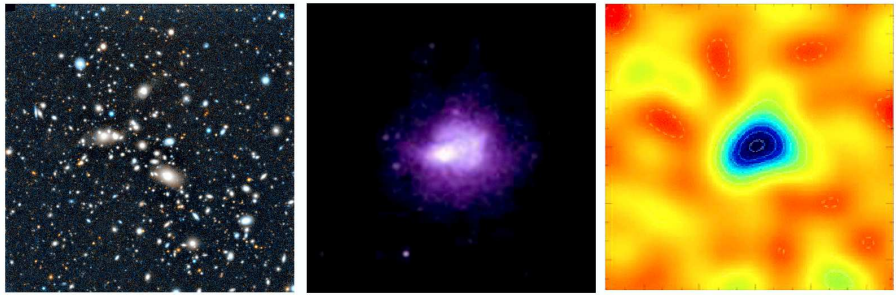}
\caption{The cluster Abell 1914 as detected in different wave-bands. 
In the optical, a galaxy cluster is simply a cluster of galaxies, 
shown as the structure in the center  of the figure on the left. In X-rays, a galaxy 
cluster is a glowing ball of electrons, emitting by thermal bremsstrahlung, with most 
of the baryons (the ordinary matter) in the form of hot ionized gas between the galaxies
(center, courtesy of Bonamente). At radio-mm wavelengths, 
the same hot gas shows up as a shadow in the Cosmic Microwave Background (CMB), 
by a mechanism known as the
Sunyaev-Zel'dovich effect. Effectively, the hot electrons give a small energy
boost to the CMB photons as the photons go through the cluster, leading to a 
distortion of the spectrum of the CMB (Courtesy of J. Carlstrom)}
\label{fig:multi}       %  
\end{figure}

Another important characteristic of clusters is that they are closed systems. In other words, 
they tend to hold their gas, unlike galaxies, 
where the gas is forced out through, for example,  supernova explosions.
X-ray studies have revealed the presence of large amounts of this intra-cluster
gas which contributes to nearly $\sim 15\%$ of the total mass of the system. 
Since it is very hot, with temperatures in between  10$^7$K and 10$^8$K, 
this plasma  emits X-rays 
in the form of thermal bremsstrahlung and makes clusters the brightest X-ray sources after 
Active Galactic Nuclei (see Fig. \ref{fig:multi}).
The total mass of the diffuse intra-cluster gas is larger than that condensed 
in its galaxy members by roughly a factor of three (e.g. \cite{marhu} even if 
this is still not enough mass to keep the galaxies in the cluster!)
In particular, since clusters are closed systems, by 
assuming that their gas mass fraction is universal, one can 
estimate the total mass density of the universe.

The inner physical properties and the dynamical evolution of bound clusters
is a major subject of scientific investigation. Nonetheless, also their large scale spatial 
distribution is a statistics of great cosmological interest.  
The space density of rich clusters is of order $10^{-5}$ Mpc$^{-3}$ so that
the typical distance between cluster centers, if they were uniformly distributed in space,
would be of order $\sim 50$ Mpc. These figures can be compared with the space density of 
{\it mean galaxies} ($10^{-2}$ Mpc$^{-3}$) and their typical separations ($\sim 5$ Mpc). 
Contrary to initial expectations (Nymann and Scott 1952) \cite{NYS} clusters themselves 
are not uniformly distributed  but  are strongly correlated in space on scales nearly 
5 times larger than those of galaxies.

By measuring the spatial abundance  of clusters and its 
time evolution, one can constrain the growth rate of large-scale 
structures, thereby placing further significant constraints on the 
coherence of our standard cosmological model and on the viability of  
alternative theories of gravity.
This motivate the search for high redshift systems, an extremely challenging 
task due to the fact that at such early epochs clusters are rare and projection effects 
hamper their identification.

\section{Identification and reconstruction methods}

From a physical point of view, a group or cluster of galaxies can be defined as a 
gravitationally bound system  having a negative  total mechanical  energy 
(evaluated in the center of mass). 
This requirement (virial equilibrium) when coupled with the geometric
constraint of sphericity implies that bound clusters must have an adimensional overdensity 
$\delta \rho/\langle \rho \rangle$ with respect to the background mass density field $\langle\rho \rangle$ 
of nearly 200. 

From an astronomical point of view, identifying clusters in terms of these dynamical 
requirements turns out to be problematic.   
The overdensity criterium is only a necessary condition.
It does not guarantee  that systems satisfying to it are 
virialized. In particular
the threshold sets a constraint on the overall distribution of matter in real space, 
not on the density contrast of the visible fraction in redshift space, i.e. the quantity 
which is directly accessible by observations.
Additionally, only three of the seven parameters (3D-positions, 
3D-velocities and masses) necessary to evaluate the energy of the system
are actually observables. These are the angular (sky) 
position of galaxies and their redshift, i.e. their distance from the observer.

As we will see, one can bypass the incomplete knowledge of the coordinates needed to identify 
a cluster  in phase space,  by introducing some additional
hypotheses concerning  the symmetries of the system or the physical properties of its members.
For example, it is fairly intuitive to characterize clusters as large and symmetrical 
conglomerates,  
the rich and  poor categories being  essentially defined by the number of red galaxies 
located within a given distance of the order of a few Megaparsec
from the center of the system (see Fig. \ref{fig:multi}). 
In essence, one aims at identifying overdensities in redshift space, luminosity and/or color space, depending on the availability of redshift information and/or photometry. It is clear, that 
the degree of objectiveness of the cluster identification algorithm fully rests on the 
robustness of these external, model-dependent, constraints. 

Given the limited number of observables and the large
uncertainties with which they are estimated, one might think 
the identification of bound structures to be quite hopeless. It is not. The situation
is saved by the statistical nature of the problem. Since clusters contains many
members and there are many clusters in the universe, one can find in 
the theory of statistics the useful theorems and tools to average out
reconstruction errors and imperfections.

With this caveat in mind, the ideal group-finding algorithm would be a  single method that can 
robustly identify and determine the membership of groups and clusters of galaxies across
a wide range of richness, mass,  redshift and surveys sampling rate.  
(As an example galaxy structures may 
range from $\sim10^{13}M_{\odot}$  to  $\sim10^{15}M_{\odot}$  in mass  and are expected to 
densely populate the universe even at such early epochs as  $z=1.5$). Additionally, 
the algorithm  should impose minimal constraints on the physical properties of the clusters, 
to avoid selection biases. If not, these biases must be properly characterized.

The optimally reconstructed  catalog of groups ought to fulfill two criteria: 
first it should be  {\it complete}, in the
sense that all the objects which fulfill the selection criteria are contained in the catalogue.
Second it should be {\it reliable} i.e. it should  not contain any object that do not fulfill
the selection criteria (the so-called false positives). More specifically,
the algorithm should be able to produce a catalog in which 
(1) all galaxies that belong 
to real groups are identified as group members, (2) no field galaxies are misidentified 
as group members, (3) all reconstructed groups are embedded within real, virialized 
dark-matter halos, (4) all real groups are identified as distinct objects \cite{ger}.

\subsection{Finding galaxy systems in 2D}

\begin{figure}
\centering
\includegraphics[height=12cm, scale=1.8]{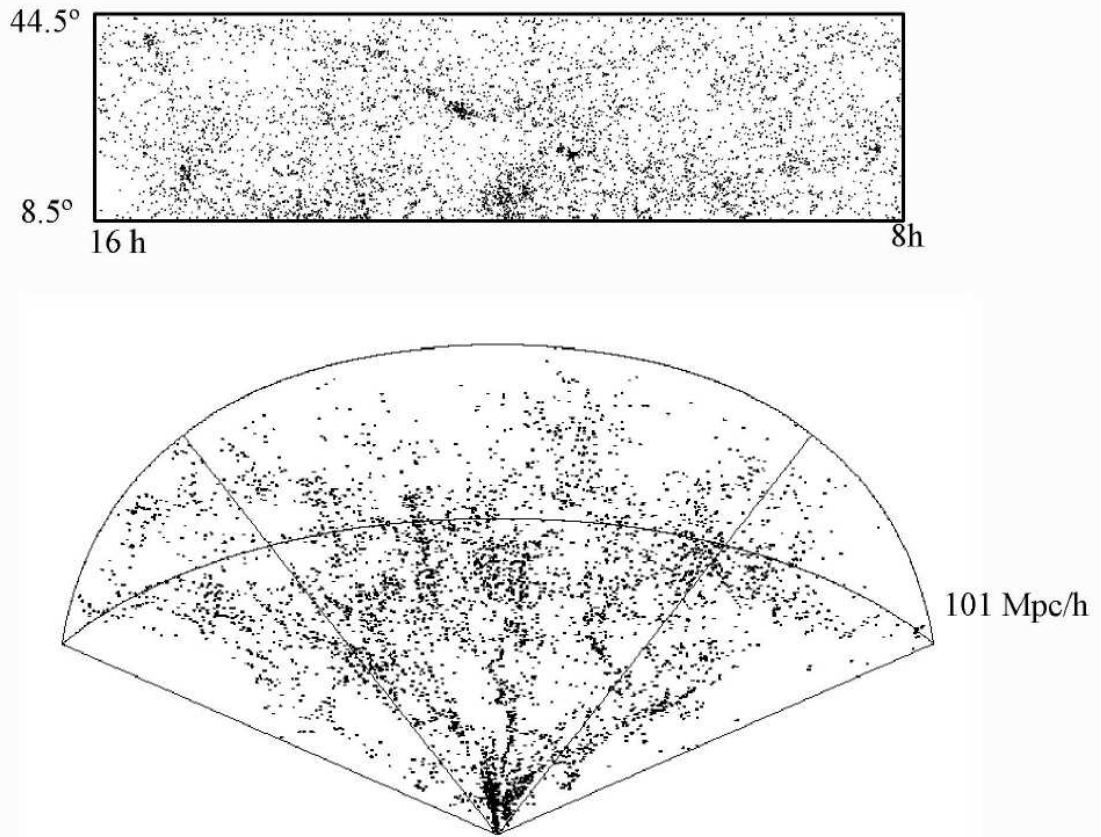}
\caption{{\it top:} 2D map of the galaxy distribution in a flux-limited survey. 
Galaxy positions are specified using angular coordinates. Clusters are easily 
recognized in this sky map as galaxy overdensities on an otherwise smooth background. 
A difficult task is to get rid of fore- and back-ground objects which do not belong
to clusters, or to identify weaker peaks usually associated to poor or distant 
clusters (only the most luminous members of which show up in a flux-limited galaxy survey)
{\em Bottom:} the same galaxy distribution  is shown in 3D space. Galaxy distances are 
inferred by interpreting the redshift, i.e. the relative shift of light wavelengths between emission and absorption, as due to the cosmological expansion of space. The redshift, which is an intrinsically kinematic observable,  is  also sensitive to non-cosmological 
Doppler contributions. Because of this spurious contamination (which is induced by the local motions of galaxies inside the gravitational potential well of a cluster)
galaxies belonging to a gravitationally bound structure such as clusters 
appear smeared out along the  observer line of sight in 3D maps,  the so-called 
"Fingers-of-God" phenomenon. (Courtesy of V. Martinez)}.
\label{fig:distortions}       %  
\end{figure}

Obtaining spectral information (distance measurements) is 
extremely demanding in terms of instrumental complexity and  observing time.  
Therefore, it is critical to develop efficient group searching strategies 
able to detect intrinsically three-dimensional clusters
in bi-dimensional data such as photographic plates, CCD images, catalogs of 
angular positions of the galaxies, etc (see top panel of figure \ref{fig:distortions}).

One of the most enduring legacies George Abell has left to astronomy is the northern 
sky catalog of galaxy systems that he compiled in the 1950's\cite{abel}. 
By means of a visual inspection 
(and the aid of a $3.5x$ magnifying lens)
Abell surveyed the then recently completed
Palomar Sky Survey photographic plates, and identified 2,712 density 
excess in the galaxy distributions. 

According to his identification algorithm, a cluster is a compact structure with
fifty or more members lying within one ``counting radius" of the cluster's center
(now called the Abell radius). Additionally, he divided the clusters into six ``richness" 
groups, depending on the number of galaxies in a given cluster that lie within the magnitude 
range $m_3$ to $m_{3+2}$ (where $m_3$ is the magnitude of the third brightest member of the density excess).  

The Abell cluster catalog is by far the most widely used catalog to date in the local universe. 
However, this ground breaking work has not stood the test of time as a valuable and efficient 
way for evaluating galaxy systems. Although the human eye is a sophisticated and efficient 
detector for galaxy clusters, it suffers from subjectivity and incompleteness, and visual 
inspection is extremely time consuming. Above all, the whole procedure 
is suitable only for rich clusters and does not apply to less massive systems characterized
by a lower density contrast with respect to the background of unrelated objects.
For cosmological studies, the major disadvantage of
such visually constructed catalogs is that it is difficult to quantify selection biases. 
In this sense the Abell catalogue is neither complete nor is it reliable. 

Only in recent  years it has become possible to use numerical algorithms 
in the search for galaxy clusters in wide field optical imaging surveys.
These modern studies required that photographic  plates  be digitized (so that the
data are in machine readable form) or that the 
data be digital in origin, coming from CCD cameras \cite{gal}. 
   
\subsubsection{Counts in cells method}

Shectman (\cite{She}) was the first to use an automated method to search for clusters
in 2D catalogs.
His box-counting technique is based on identifying local density maxima above a threshold 
value, after 
smoothing data with a weighted kernel. This algorithm
uses sliding windows which are moved across the point (galaxy) distribution 
marking the positions where  
the count rate in the central part of the window exceeds the value
 expected from the 
background determined in the outermost regions of the window.

However kernel smoothing  invariably reduces the amount of information that can be retrieved
from data. In this type of algorithms, for example, the kernel is fixed in angular size and, 
as a consequence, it does not smooth clusters at  different distances with the same efficiency, 
making its sensitivity highly redshift dependent.
The main drawbacks of the method are the introduction of a binning to determine 
the local background, which improves count statistics at the expense of spatial accuracy.

A slightly more sophisticated technique is to use an adaptive smoothing kernel 
\cite{sil}. Even the introduction of locally adjustable searching parameters, however,
does not significantly improve the performances of the technique when it is applied
to galaxy catalogs spanning a wide range of depths. Moreover,  
cluster richness is evaluated in a step separate from  detection, which further complicates the 
implementation of the method.

\subsubsection{The matched filter technique}

A substantial improvement in efficiency in finding clusters in two-dimensional optical data 
was obtained with the `matched filter' technique  of Postman et al. (\cite{post}).
This technique,  widely used in telecommunication (where it is known as the North filter)
consists in  correlating a known signal (or template), with an unknown signal to 
detect the presence of the template in the noisy background.

Postman et al. analyzed the galaxy distribution by adopting  
an a-priori cluster model to fit the data.
In particular,  by assuming a specific  cluster radial density profile 
and a cluster luminosity  function they  constructed a matched filter in 
position and magnitude space.
In practice, the 2D sky image is convolved with the assumed cluster model;
all structures that resemble the filter will be filtered as little as possible, 
but structures that are different will get attenuated.
This method, therefore, converts a  2D sky image into a 2D likelihood map.
The intensity of the resulting map approximate the probability of each point being 
the center of a cluster. 

Postman et al. realized that 
using magnitudes, rather than simply 
searching for spatial density enhancements, 
suppresses false detections that occur by chance projection. 
The matched filter is a very powerful cluster detection technique. It can handle deep 
surveys spanning a large redshift range, and provides redshift and richness measures 
as an innate part of the procedure.
However, the main drawback of the method is that 
clusters which have properties inconsistent with the input functions 
will be detected at lower likelihood, if at all. In particular 
it can miss clusters that are not symmetric or whose density distribution  
differ significantly from the model profile. Therefore this algorithm does not 
performs optimally  over 
large redshift baselines where {\it a-priori} unknown evolutionary effects 
are expected to modify significantly the simple representations used as input 
in the local universe (see the discussion by \cite{Gal}).
Nevertheless, this remains one of the best cluster detection techniques for cluster
detection in moderately deep surveys.

\subsubsection{Photometric techniques}

Systematic searches for clusters using galaxies as signposts for detection 
are mostly based on identifying a class of special 
objects supposed to live  preferentially in high density environments. 

One approach consists in looking for clusters directly in the sky.
The strategy is to assume that luminous Active Galactic Nuclei (AGN) 
are found preferentially in high density regions. This is 
actually expected in models of galaxy formation.
If the redshift of the AGN is known, one can use 
opportunely selected  narrow-band filters to survey the regions
around the AGN in quest of possible neighbors. In other words, by 
using a purely photometric technique one looks for 
the existence of a density excess of objects at the same distance of the AGN. 

Another approach exploits the fact 
that late-type galaxies are the dominant population in the field, whereas 
early-types are
preferentially seen in high-density regions (see Fig. \ref{fig:md}).
Therefore, any technique that can  eliminate field (i.e., late-type) galaxies on the 
basis of some simple observable parameter will enhance the contrast of galaxy clusters 
relative to the background.  For example, a popular 
algorithm  uses the color-magnitude relation as a filter (see Fig. \ref{fig:cm}) to select
possible cluster members (the {\it Red Sequence} method \cite{YGL}).

The implementation of these  methods requires that images of the same 
sky region be available in two different optical bands. In particular since,  
the elliptical ridge (red sequence) is such a strong indicator of a cluster's presence, 
this technique  can be used to detect 
clusters to high redshifts ($z \sim 1$) with comparatively shallow imaging, 
if an optimal set of photometric bands is chosen.
Moreover the redshift evolution of the red cluster sequence is so precisely characterized, 
that from the 
color-magnitude diagram of a cluster alone, its redshift can be estimated, whereby a typical 
accuracy of $dz \sim 0.1$ is achieved.

While using a color-magnitude relation to identify possible cluster members could 
enhance  the identification success rate, it is also likely to suffer from selection biases, 
such as missing systems  with significant blue populations of galaxies, namely, the 
Butcher-Oemler (\cite{BO}) clusters. As a consequence, one should not overlook the 
possibility that the resulting catalogs do not provide a complete census of the cluster 
population. 
 
Recent methods have been developed to minimize the impact of these selection biases.
For example the MaxBCG \cite{han,koe} or the K2 \cite{than} methods are specifically 
designed and tailored to detect galaxy clusters in 
multicolor images. These identification algorithms look for `distinctive signatures' 
such as the simultaneous galaxy density enhancements in both colors and position spaces, 
as well as the eventual presence of a bright galaxy in the targeted region.

Another class of methods is intermediate between 2D and 3D cluster finders.  
For example, Adami et al. 2010 \cite{adami3} used an adaptive kernel technique 
to reconstruct density maxima from photometric
redshifts (i.e. redshifts estimated on the basis of galaxy colors as opposed to
galaxy spectra). This approach has the drawback of being ``{\it data demanding}",  
in that it requires a multi passband 2D survey in at least 5 photometric bands. 
Nonetheless, at variance with purely 2D methods, the method opens up the possibility 
of searching  for very distant  ($z>1$)  clusters.

\subsubsection{Non-optical techniques}

The various searching strategies we have described try to eliminate some of the 
subjective criteria and assumptions of past attemps, particularly detection by eye. 
Notwithstanding, a common problem
of all these algorithms is that the recovered cluster samples are statistically complete only 
near the upper tail of the mass distribution. In effect, these methods identify only those rich 
aggregates that are most conspicuous. However, less extreme systems such as galaxy groups, which 
contain most of the luminosity, and presumably mass, of the universe, may be more useful probes 
of the 
large-scale structure. Additionally, in the absence of  information about galaxy distances,
projection effects are a serious and difficult to quantify issue.
In photometric surveys, the increased depth necessary for high-redshift studies increases the
overall number density of objects, thereby increasing the problems of foreground and background
contamination and projection effects (although photometric techniques for estimating redshifts 
can mitigate these difficulties).

Since identification biases, essentially due to foreground and background contamination, 
plague any optically selected cluster sample (and have often translated into flawed
scientific conclusions) alternative approaches to identify clusters in 2D have been
investigated: X-ray emission from hot
 intra-cluster gas \cite{ros,pierre,fino}, 
 strong lensing induced by the cluster gravitational  potential \cite{kneib2003,limousin2009,richard2010}, cosmic shear due to weak gravitational
 lensing \cite{ref, gavazzi2009}, the Sunyaev-Zel'dovich (SZ) effect in the cosmic
 microwave background \cite{car, voit}.   
 However, most of the methods used for local
 studies have only limited effectiveness at high redshift. The apparent surface brightness
 of X-ray clusters dims as $(1 + z)^{-4}$, making only the richest clusters visible at high
 redshift. The cross section for gravitational lensing falls rapidly at high redshifts,
 making weak-lensing detection of distant clusters difficult for all but the most massive
 objects. The SZ effect is very promising, since it is entirely independent of redshift, 
 but it also suffers from confusion limits and projection effects, and, in any case, large
 surveys of SZ clusters, such as those promised by the South Pole Telescope \cite{Staniszewski} 
 or the Planck satellite \cite{bartlett}, are yet in a preliminary stage.

\begin{figure}
\centering
\includegraphics[height=9cm, angle=-90]{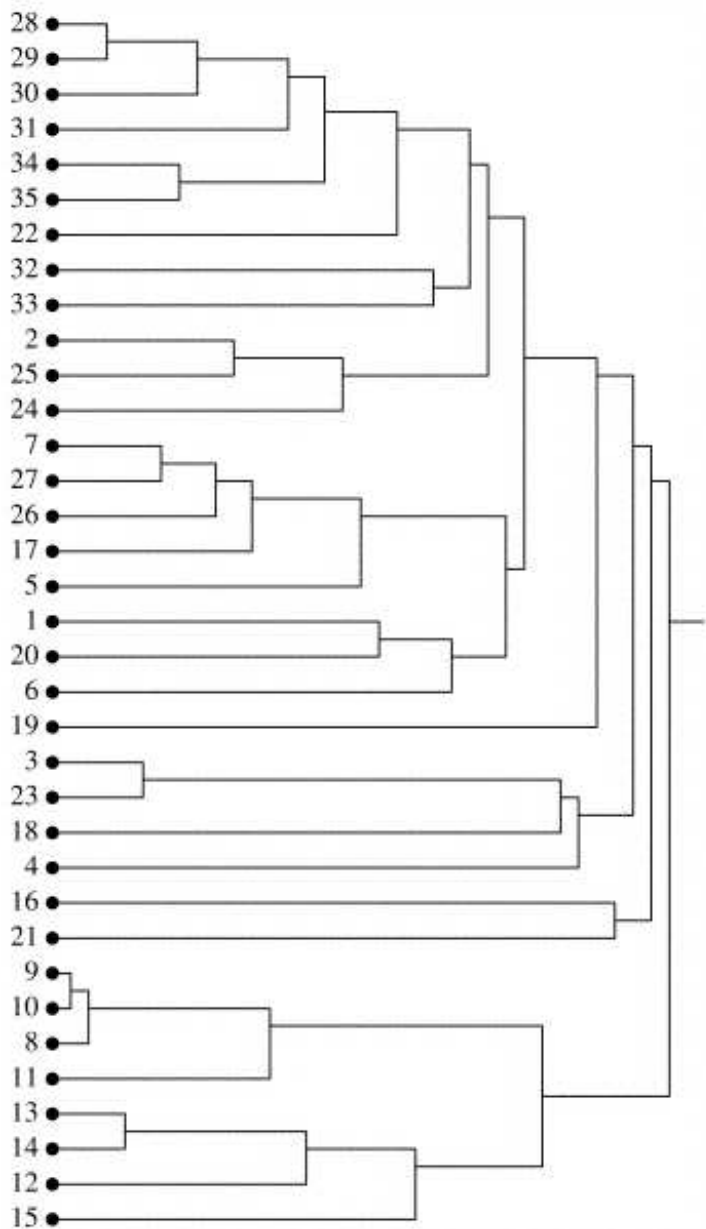}
\caption{Dendogram showing how a set of galaxies merge and form a hierarchy.    
One starts with all galaxies of the sample 
as separate units and links the units successively in order of decreasing affinity 
until there is only one unit that encompasses the ensemble.
The vertical axis measures the decreasing strength of the affinity parameter
according to which galaxies are linked
The limiting value of the affinity parameter 
below which one define clustered and isolated units must be
fixed using independent physical considerations
(Taken from \cite{gaz}.)}
\label{fig:dendo}       %  
\end{figure}

\subsection{Finding galaxy systems in 3D}

The first sizeable sample of groups detected in redshift space was presented in 1983 
by Geller \& Huchra \cite{GH}, who found 176 groups of three or more galaxies 
in the CfA galaxy redshift survey at 
redshifts $z < 0.03$. Recently, Yang et al. \cite{yang} identified groups within the 4th  data 
release of the Sloan Digital Sky Survey (SDSS). Their catalog extends to $z < 0.25$ covers
one tenth of the sky  and constitutes the largest currently available catalog of galaxy groups,
containing $\sim 3\cdot10^5$ groups with two or more members. 

Despite the additional dimension, identifying in an unbiased way groups and clusters in 
redshift space is  a formidable task. 
The most obvious and well-known complication is redshift-space
distortions: the orbital motions of galaxies in virialized
groups cause the observed group members to appear spread out
along the line of sight (the Fingers-of-God effect), while coherent
infall of outside galaxies into existing groups and clusters reduces
their separation from group centers in the redshift direction (the
Kaiser effect \cite{kaiser}). Both of these effects confuse group membership
by intermingling group members with other nearby galaxies (see Fig. \ref{fig:distortions}).

Since it is impossible to separate the peculiar velocity field from
the Hubble flow without an absolute distance measure \cite{dek,mari98}, this confusion
can never be fully overcome, and it will be a significant
source of error in any group-finding program in redshift space. 

A second complication arises from incomplete sampling of the galaxy population. 
No modern galaxy redshift survey can succeed in measuring a redshift for every 
target galaxy, and an incomplete 
galaxy sampling rate always leads to errors in the reconstructed catalog of groups 
and clusters even without redshift-space distortions. 

Moreover, surveys conducted over a broad redshift range present 
their own impediments to group finding. The major problem is that distant galaxies appear 
fainter than nearby galaxies. At high redshift one can  probe only relatively rare, 
luminous galaxies, so only a small fraction of a given group's members will meet a given 
selection criteria.

\subsubsection{The hierarchical method}

In the astronomical community the most widely used objective 3D group-finding algorithms are 
the hierarchical and the percolation (friends-of-friends) algorithms.
In the hierarchical  clustering method, first introduced by Materne (\cite{Mat}), one defines an 
affinity parameter between the galaxies, which controls the grouping operation.  There are 
several possible choices for the grouping parameter. The most widely adopted choice is to use 
the product of galaxy luminosities divided by separation squared, which is a proxy of the
 gravitational force between galaxies i and j. Then one starts with all galaxies of the sample 
as separate units and links the galaxies successively in order of decreasing affinity 
until there is only
 one unit that encompasses the ensemble. A hierarchical sequence of units organized by
 decreasing affinity is the result of this method. The merging of a galaxy into a given unit
 involves the consideration of the whole unit and not just of the last object merged into the
 unit. Another merit of this method is the easy visualization of the whole merging procedure
 under the form of a hierarchical arborescence, the dendrogram (see Fig. \ref{fig:dendo}).

  Customarily, it is believed that the Hierarchical  method has the practical drawback of
 requiring a very long calculation time (e.g., in comparison with the percolation method). 
Paying attention to this problem, Giuricin et al. (\cite{giu}) 
have managed to considerably speed 
up the hierarchical code by using numerical tricks. In this way, one can run this code nearly 
as fast as the percolation algorithm. For example the C programming language allows to use 
techniques of sparse matrix (i.e., with most elements equal to zero) in a natural way,
through a data structure based on pointers. In this way, for each pair of  galaxies, the 
affinity parameter, is not stored in memory and is not exactly calculated but replaced with
 zero if its value is smaller than a preselected limit. The maximum value of this parameter
 is searched only for the few pairs for which the parameter values are greater than this limit.
 Then the limit is gradually lowered in the following steps until the dendrogram is completed.

The main drawback of this procedure is that the threshold value where one cuts the hierarchy 
and below which one accepts the clusters as real is not supplied by the method, 
and must be fixed arbitrarily. 
A common choice is to cut the hierarchy according to the luminosity density 
or  number density of the unity. This is a parameter which is highly sensitive to 
redshift distortions (see Fig. \ref{fig:distortions}). 
In particular rich structures can have very low apparent densities in redshift space because 
their members are scattered away along the line-of-sight by the cluster potential.
Clearly by lowering the density selection  threshold, 
one could  in principle  recover  these real structures but at the price of  
contaminating the  catalog with lots of spurious, artificial  structures.

\subsubsection{The percolation method}

The percolation or friends-of-friends (FOF) algorithm \cite{HG},  being easier 
to implement than the Hierarchical one, has been the most widely used method of group
 identification in 3D. Moreover,  since this technique has a natural theoretical 
motivation in the context  of current model of galaxy formation, 
it has been extensively applied to detect overdensities not only in redshift surveys 
but also in N-body simulations of the large scale structure.

\begin{figure}
\centering
\includegraphics[height=11cm, scale=8.8]{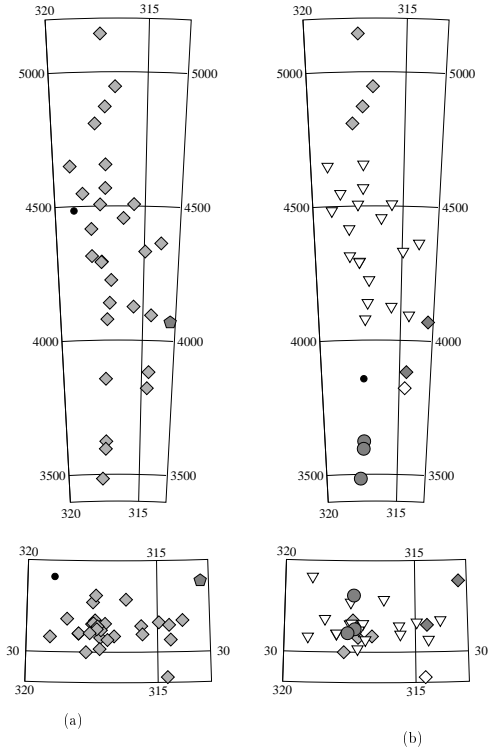}
\caption{
Comparison of the performances of the FOF and Hierarchical methods
in identifying cluster members in high density regions.
Galaxies in the region of the Abell 3574 clusters are shown in projection
along the line-of-sight (upper panels) and on the sky (lower panels).
Similar symbols are use to label galaxies identified as members of the same
group by the percolation method (left panel) and the hierarchical method (right panel)
(Taken from \cite{mari}.)}
\label{fig:comp1}       %  
\end{figure}

\begin{figure}
\centering
\includegraphics[height=11cm, scale=8.8]{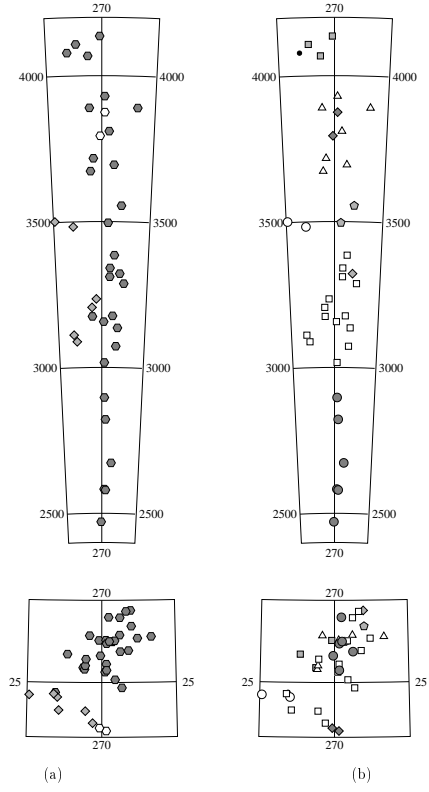}
\caption{
Same as in Fig. \ref{fig:comp1} but in the region of the the Hydra overdensity
(Taken from \cite{mari}.)}
\label{fig:comp2}       %  
\end{figure}

Unlike the Hierarchical algorithm, this technique does not
rely on any {\em a-priori} assumption about the geometrical shape of groups.
All pairs linked by a common ``friend" form a group if the 
number overdensity contrast  exceed an arbitrarily fixed critical threshold $\delta n/ n$. 

We here present more in detail the FOF algorithm adopted by Eke et al. \cite{eke}.
Consider two galaxies $i$ and $j$ with comoving 
distances from the observer $d_i$ and $d_j$ respectively. 
These two galaxies are assigned to the same group if their 
angular separation $\varphi_{ij}$ satisfies

\begin{eqnarray}
\varphi_{ij} \leq \frac{1}{2} \left(\frac{r_{\perp,i}}{d_i}+\frac{r_{\perp,j}}{d_j}\right)
\end{eqnarray}

and, simultaneously, the difference between their distances satisfies
\begin{eqnarray}
\left|d_i-d_j\right| \leq \frac{r_{\parallel,i}+r_{\parallel,j}}{2}.
\end{eqnarray}

\noindent where $r_{\perp}$ and $r_{\parallel}$ are the comoving linking 
lengths perpendicular and parallel to the line of sight.
In order to take into account the decrease of the magnitude range of 
the luminosity function  sampled at increasing distance in flux limited surveys, 
the projected link parameter, 
$r_{\perp}$ and $r_{\parallel}$
are in general suitably increased 
with increasing distance. By scaling them,  one keeps the number 
density enhancement $\delta n/ \langle n \rangle$ constant. However, the scaling prescriptions, 
besides being  somewhat arbitrary, introduces serious biases in the reconstructed
group catalog. Eke et al. adopted the following scheme

\begin{eqnarray}
r_{\perp} &=& \min \left[r_{\rm max}(1+z),\frac{b}{\langle n \rangle}^{1/3}\right]\\
r_{\parallel} &=& R\; r_{\perp},
\end{eqnarray}

\noindent arguing that scaling both $r_\perp$ and $r_\parallel$ with $\langle n \rangle^{-1/3}$ 
will compensate for the magnitude limit and lead to groups of similar shape and 
overdensity throughout the survey. 

The FOF algorithm  has three free parameters: the linking length $b$, 
the maximum perpendicular linking length in physical coordinates $r_{\rm max}$, and 
the ratio between the 
linking length along and perpendicular to the line of sight $R$. 
The free parameter $r_{\rm max}$ has been introduced to avoid unphysically large values 
for $r_\perp$ at high redshifts 
where the galaxy distribution is sampled very sparsely. Since $r_{\rm max}$ is 
measured in physical 
coordinates, $r_{\rm max}(1+z)$ is the maximal comoving linking length perpendicular to the line of sight. 
The free parameter $R$ allows $r_\parallel$ to be larger than $r_\perp$ taking into account the 
elongation of groups along the line of sight due to the Fingers-of-God effect.
Finally, the linking parameter $b$ can be constrained, in real space,  
by modelling a cluster as a bound structure resulting from the 
non-linear gravitational collapse of a spherical perturbation. 
Because of this, the percolation algorithm is a natural method
for identifying virialized structures in the absence of redshift-space distortions
and is largely used to recover groups and clusters in real-space N-Body numerical
simulations.

In figures \ref{fig:comp1} and \ref{fig:comp2}   two characteristic high 
density regions (Abell 3574 and Hydra clusters) of the universe are shown
together with the cluster identified by two different group-finding algorithms:
the FOF  and Hierarchical methods.
By inspecting them one can qualitatively contrast  the 
performances of these two methods. The hierarchical algorithm, 
selecting preferentially systems with a spherical symmetry, fails to 
recover massive clusters with prominent {\em Fingers of God}, whose members are
better identified  using the percolation algorithm. 

These standard cluster-finding algorithms are less than optimal for many reasons. For instance, 
the searching window at the heart of  FOF techniques is insensitive to local variations 
in the density of points. Assigned cluster membership therefore depends on the scale of the 
adopted linking length and not the distribution of galaxies alone, violating the dictum to 
"let the data speak for themselves". In fact, both the hierarchical and the
 percolation methods require prior knowledge and/or user-fixed parameters to produce their 
best results. Density thresholds, linking-length parameter scaling laws, galaxy selection 
functions, etc., must all be set in advance. The preprocessing and/or trial-and-error tests
 required to tune these algorithms for a particular data set are extremely inefficient and 
may even lead to systematic differences among different applications of the same technique.

     It is well known that the performance of the standard FOF algorithm across a wide range 
of density enhancements is not uniform. A generous linking length is 
preferred in studies that aim to identify high velocity dispersion systems. On the other hand,
studies of loose associations require short velocity links, but this can result in a bias 
toward low velocity dispersion measurements. In general, the velocity dispersion of systems 
identified with the FOF algorithm is $\sim 30\%$ higher than the velocity dispersion of groups 
identified in the same galaxy catalog by the hierarchical method (\cite{giu}). 
To further complicate matters, clusters identified with one method may not be detected by 
the other.

Given the weaknesses/failings of traditional cluster
identification methods, which are likely to only be worse at high
redshift (a domain which is progressively conquered by modern observational
campaigns), there have been 
various attempt to explore new detection strategies.
Kepner et al. (\cite{kepner}) generalized in 3D 
the matched filter   algorithm. This new adaptive code  
identifies clusters by adding   halos  
to a synthesized background mass density and computing the
maximum-likelihood mass density. The Sloan Digital Sky Survey team  
has introduced a group-finding algorithm called C4 (\cite{nichol}), which
searches for clustered galaxies in a seven-dimensional space,
including the usual three redshift-space dimensions and four
photometric colors, on the principle that galaxy clusters should
contain a population of galaxies with similar observed colors.     
In quest of the optimal algorithm, various researchers have also explored  
the possibility of using the geometries of  Voronoi and Delaunay.

\section{Voronoi based group-finding tools}

The Voronoi tessellation made his ``debut" in cosmology as an useful theoretical framework 
for interpreting the clustering of galaxies  (\cite{IW,WI})
\footnote{In fact, what became known
as the Voronoi diagram was first suggested in an astronomical context for a problem 
somewhat similar to that investigated here. In his treatment of cosmic fragmentation
Le monde, ou, Traité de la lumière, posthumously published in 1664, Réne Descartes 
used Voronoi-like methods to model the spatial distribution and the relative influence 
of solar system bodies (see Fig. 7 in \cite{obs}).}.
Since then, it has been used  as a valuable method for investigating several issues concerning
the large-scale structure of the universe (see \cite{phdvw} and the review of
Van de Weygaert in this volume.)

The Voronoi partition of a space into minimally sized convex polytopes
-- the three-dimensional analogue of Dirichlet tessellation or
determination of Thiessen polygons  -- provides a
natural way to find cluster centers (peaks in the galaxy density
field).  The volume inside each polyhedron is inversely 
proportional to the packing efficiency of its seed; a large cell volume 
indicates that its seed is comparatively isolated.
While other methods  estimate the galaxy density  field  by smoothing
the distribution of data points with an {\em a priori} physical model,
window profile or binning strategy,
the  Voronoi diagram provides  a density  estimator  that is
asymptotically local: the density measured at a position
${\bf x}$ is determined completely by the positions of the neighboring
data points, while the influence of distant points vanishes.

The Delaunay complex, the
simultaneously-determined dual of the Voronoi diagram, implicitly
contains vast amounts of proximity information. It  yields a natural
measurement of inter-galaxy scale lengths while remaining linearly proportional
in size to the dataset. By providing a natural
linking structure for a set of objects, the Delaunay triangulation allows
to reconstruct the members of a galaxy cluster.

\begin{figure}
\centering
\includegraphics[height=11cm]{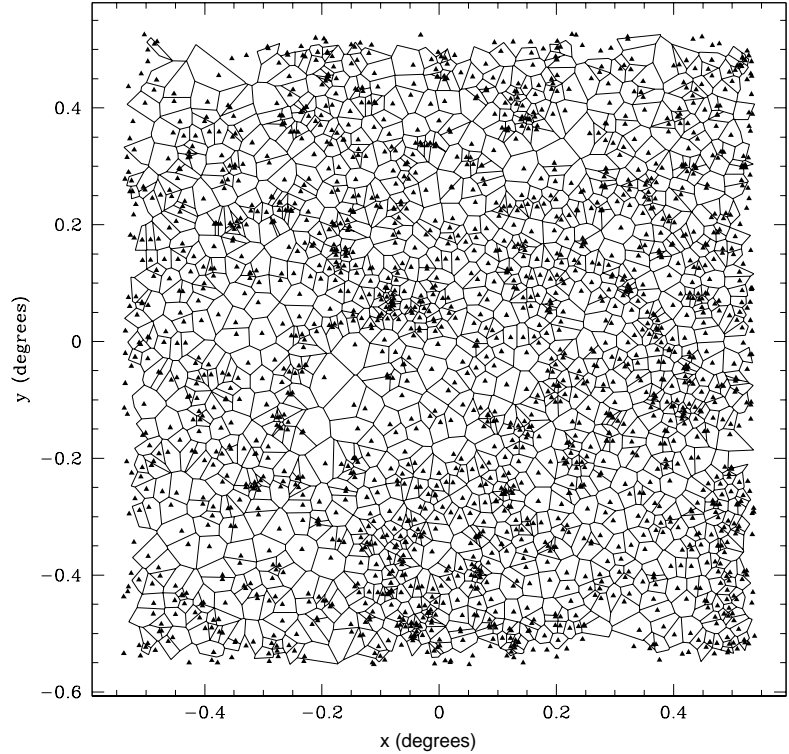}
\caption{Bidimensional Voronoi tessellation of the distribution of galaxies 
in a particular sky region. 
Each triangle represents a galaxy surrounded by its associated Voronoi cell 
(indicated by the polyhedrals). 
The area of this cell is interpreted as the effective area a galaxy occupies in the plane. 
The inverse of this area gives the local density at that point. 
Galaxy clusters are identified by high density regions, composed of small adjacent cells, 
i.e. , cells small enough to give a density value higher than the chosen density threshold
(Taken from \cite{ram}.)}
\label{fig:2dv}       %  
\end{figure}

In what follows I give an overview of various Voronoi-based  
group-finding methods that have been developed in an astronomical context. I will first 
present algorithms that have been conceived for cluster identification in 2D imaging surveys.
Then, I will  discuss in greater detail the Voronoi-Delaunay Method (VDM) proposed by Marinoni
et al. (\cite{mio}) for identifying structures in 3D redshift surveys.

\subsection{2D Voronoi algorithms}

Brown (1965) and  Ord (1978) were the first to suggest the
use of Voronoi cell volumes as density measures.
The algorithm of Ebeling \& Wiedenmann (\cite{ew}), in particular, 
was the first Voronoi based  peak finder method in the astronomical literature.
It was introduced as an optimal way to detect overdensities in X-ray photon counters and 
thus identify astronomical X-ray sources.
Since then, Voronoi based density estimators have rapidly grown in popularity (and 
sophistication) and have been applied to investigate a large class of  astronomical 
phenomena (see \cite{Schaap00,scha,WeySchaap09}).

Ebeling \& Wiedenmann partitioned the detector surface using 2D Voronoi
cells and defined as overdensity regions those composed by
adjacent Voronoi cells with a density higher than the chosen threshold. 
They used a rigorous statistical approach  for setting the detection criteria. 
An empirical distribution function describing 
randomly positioned points following Poissonian statistics, has been
proposed by Kiang (\cite{k}):
\begin{equation} 
dp(\tilde{a})=\frac{4^4}{\Gamma(4)}\tilde{a}^3 e^{-4\tilde{a}}d\tilde{a},
\end{equation} 
where $\tilde{a}\equiv a/\langle a \rangle$ is the cell area in units of the average
cell area $\langle a \rangle$. 
The idea of Ebeling \& Wiedenmann  was to estimate the background noise
by fitting the Kiang  cumulative distribution 
to the empirical cumulative distribution resulting from the data points in the region
of low density which is not affected by the presence of structures
($\tilde{\rho}\equiv \rho/\langle f \rangle\,\le 0.8$). 
In this way they were able to 
minimize the contamination of spurious detections.

This algorithm can be used quite generally to find any kind of structures embedded in a noisy 
background field. Ramella et al (\cite{ram})  in particular  used it to identify galaxy clusters 
in bi-dimensional sky images as significant density
fluctuations above the background (see Fig. \ref{fig:2dv}). They noted that the procedure, 
being completely non-parametric, is particularly sensitive to both symmetric and 
elongated and/or irregular clusters. 
Sampling of the density distribution is just the first step in a cluster detection procedure, 
followed by location of the density peaks that fulfil the criteria for a galaxy cluster 
The simplest approach is to select objects with a certain contrast above the background.
Notwithstanding, the choice of the appropriate threshold is not trivial.
With increasing threshold, the detection rate of the real clusters is 
declining, but also the relative number of spurious clusters detected is dropping. 
Moreover, in the outer regions of the clusters, where the number density is gradually dropping, 
the detection of the member galaxies is strongly dependent on the assumed threshold. 
Because of this, more refined methods, instead of sharp density thresholds, 
use a Maximum Likelihood  techniques to increase the identification efficiency and 
better delineate the boundaries of the Voronoi selected clusters in 2D (see 
method and discussion by \cite{soc}).

The false/positive detection rate is hard, if not impossible, to quantify, without running an 
algorithm on a catalog with extensive spectroscopy. Anyway, one can compare the relative 
performances of different cluster-finding codes by running them on the 
same 2D photometric catalog. 
By running their Voronoi Galaxy Cluster Finder (VGCF)  on the same data
(the Palomar Distant Cluster Survey (PDCS)) processed by Postman et al. with 
their matched filter method, Ramella et al. were able to identify 37 clusters. Of 
these clusters, 12 are VGCF counterparts of the 13 PDCS clusters
detected at the 3$\sigma$ level. Of the remaining 25 systems, 2 are PDCS clusters
with confidence level $<3\sigma$.
According to Ramella et al. inspection of the 23 new VGCF clusters indicates 
that several of these clusters may have been missed by the matched filter algorithm for one
or more of the following reasons: a) they are very poor, b) they are
extremely elongated, c) they lie too close to a rich and/or low redshift cluster.

\begin{figure}
\centering
\includegraphics[height=9.5cm]{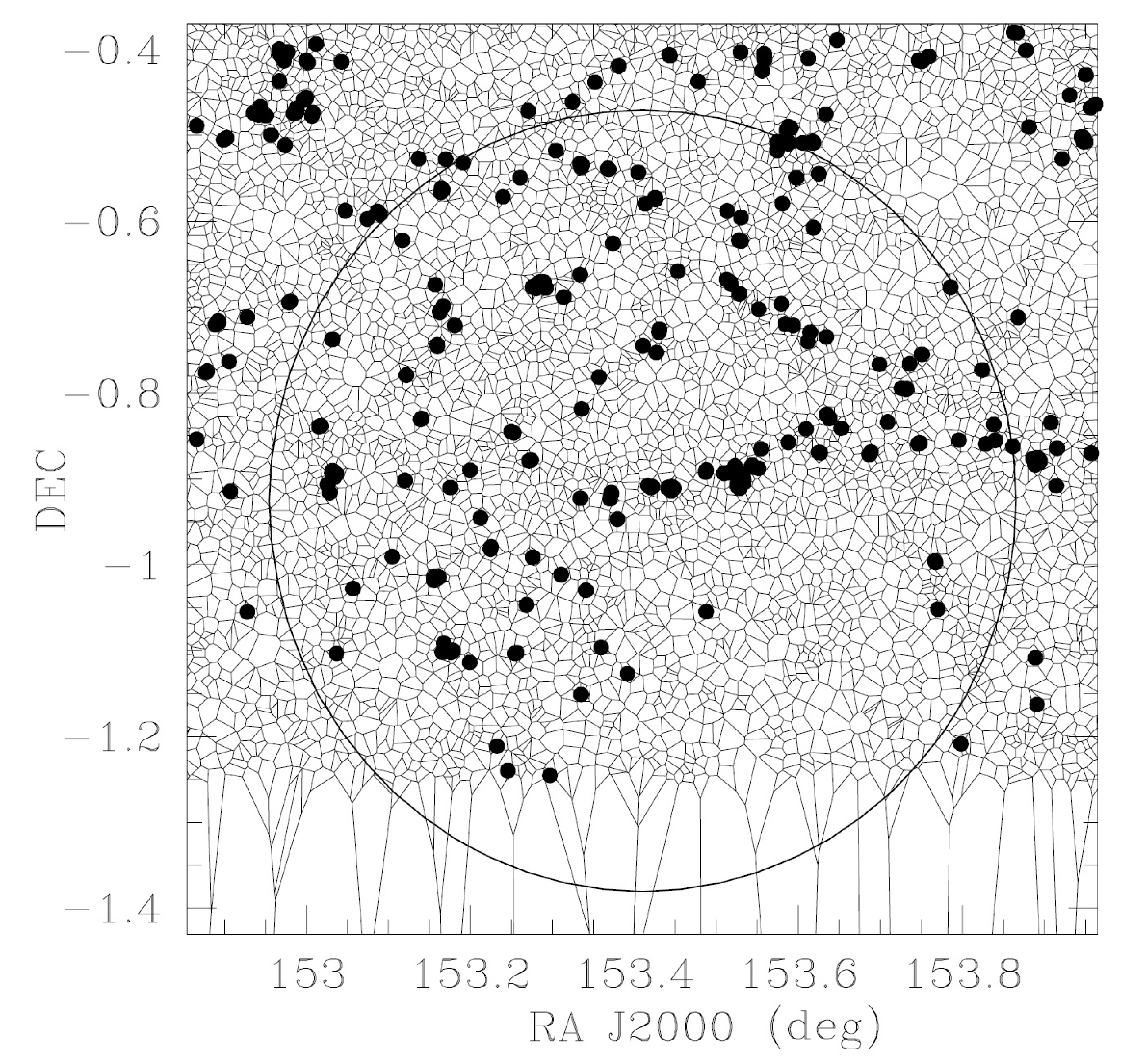}
\includegraphics[height=9.5cm]{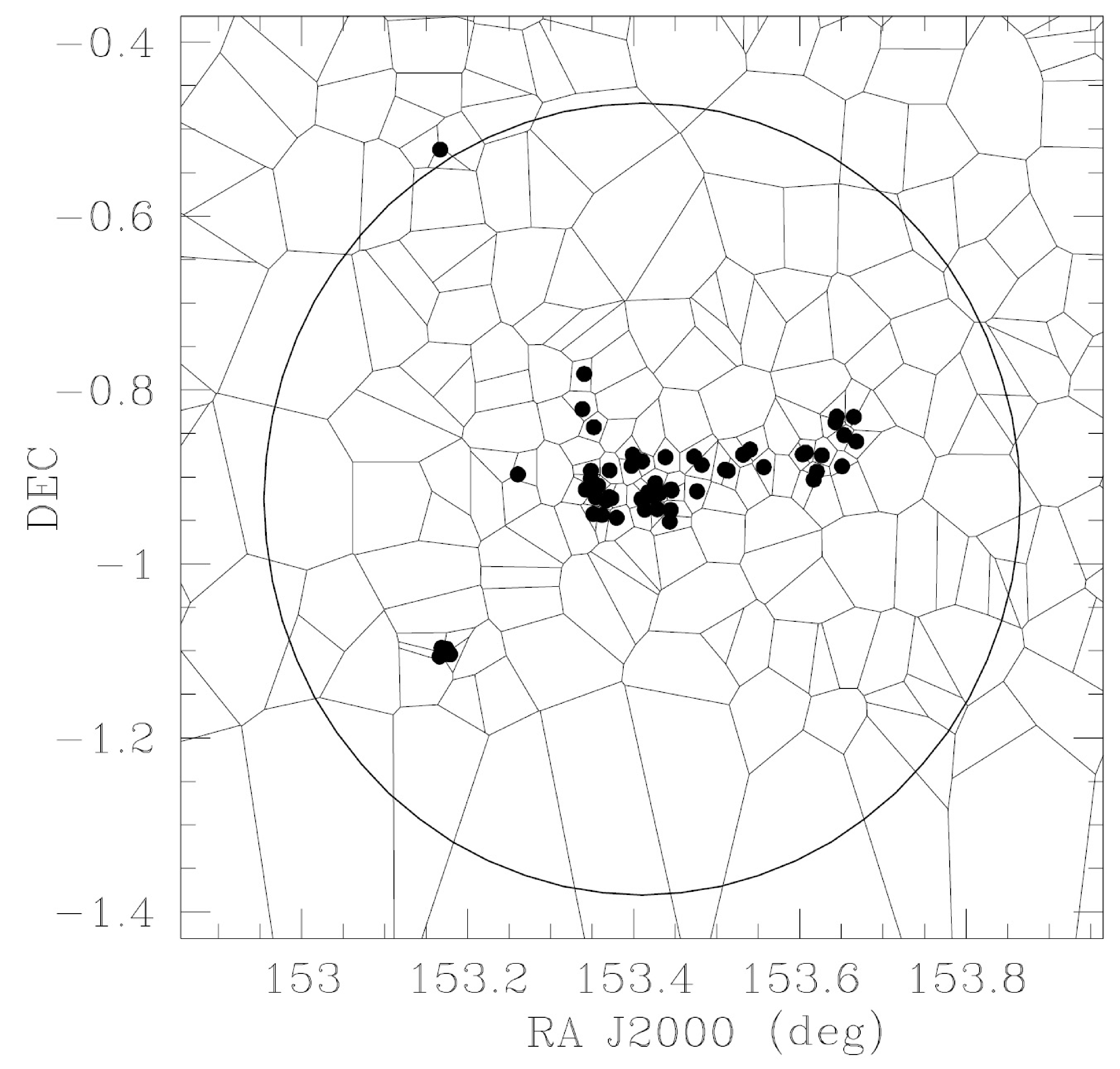}
\end{figure}

\begin{figure}
\caption{
{\it Top:} Example of Voronoi tessellation executed on the galaxy distribution 
around Abell cluster 957.
Each cell encloses one galaxy. The data presented here has a lower boundary in declination, 
which is why the Voronoi tessellation seems to diverge below. 
The  filled circles mark galaxies satisfying the selection criteria,  
$\delta (\bar{a}-a)/\bar{a} \rangle3$ where $a$ is the area of a Voronoi cell and $\bar{a}$ 
its average value. No significant overdensity of the filled circles around Abell 957 appears
when using the entire distribution of galaxies.
{\em Bottom:} same as above but the Voronoi tessellation is evaluated only on the galaxies that 
satisfy the color-magnitude criteria used in the VTT. Unlike above the cluster is now strikingly 
enhanced by the filled circles, which denote galaxies with $\delta >3$.
(Taken from \cite{kim}.)}
\label{fig:kv}       %  
\end{figure}

An upgraded version of a Voronoi based algorithm to identify galaxy clusters in 
2D images was proposed by Kim et al. (\cite{kim}). 
In order to increase the detection performance, 
their recipe for the Voronoi Tessellation Technique (VTT) uses a-priori knowledge of 
some characteristics of the cluster member 
galaxies, namely, the characteristic ridge in the color-magnitude 
diagram (see Fig. \ref{fig:cm}).  
By applying  a  filter in color-magnitude space 
they reduce the contamination of projection effects  
and  greatly enhance the cluster members contrast relative to the background.

Once the color-magnitude diagram has been used to preselect potential 
cluster members 
Kim et al. apply the   Voronoi tessellation on the resulting distribution of galaxies 
and  select only cells whose area satisfy an  empirically defined boundary condition.
In practice, potential candidates are identified by requiring a minimum number N of 
galaxies having overdensities $\delta$  greater than some threshold $\delta_c$, 
within a radius of fixed size. The top panel of figure \ref{fig:kv} 
shows Voronoi tessellation executed on all galaxies in the region of Abell 957, 
whereas the bottom one shows only those galaxies that satisfy the color-magnitude cut.
One can immediately appreciate the remarkable enhancement of the cluster, with respect
to the previous case.

By using simulations to compare their algorithm against a matched-filter method,  
they found that the matched-filter algorithm outperforms the VTT when using a background 
that is uniform, but it is more sensitive to the presence of a nonuniform galaxy background 
than is the VTT. This last method has also a better overall 
false-positive rate compared with the MF.

Lopes et al. (\cite{lopes}) applied both the VGCF and the adaptive kernel
techniques to identify clusters  over the 2700 square degrees of the digitized Second 
Palomar Observatory Sky Survey (DPOSS).  The comparison  of catalogs generated by these  
different techniques is not a 
straightforward task. As they underline, even  
clusters that are identified  by various techniques 
might have large differences in the measurement of properties such as richness, 
or projected density profiles. Notwithstanding, by contrasting  both  
algorithms  they find  that the Voronoi algorithm performs better for poor, 
nearby clusters, while the adaptive kernel goes deeper and detects  richer systems.
Anyway, they conclude that in order to obtain more complete and unbiased cluster catalogs 
it is imperative to minimize contamination by random fluctuations in the galaxy distribution 
and chance alignments of galaxy groups by complementing geometrical tools 
with multicolor photometric information such as the  color-magnitude diagram.

In this spirit, Van Breukelen et al. (\cite{van})
developed  an hybrid  cluster-finding algorithm based on a combination of the Voronoi 
tessellation  on 2D images and Friends-Of-Friends methods applied to photometric estimates of the 
galaxy redshift.  They use Montecarlo simulations to  assign a reliability factor F to each
cluster and then  cross-correlate the cluster candidates output by the Voronoi and FOF methods 
by taking all cluster candidates that are detected in both with F greater then a fixed 
threshold.

A common feature of all the previous
algorithms which make use of the Voronoi
partition to identify cluster candidates in 2D, is that cluster members identification 
is usually carried out using additional and independent methods. For example, the 
most popular technique consists in selecting as clusters members those galaxies 
which lie within fixed-radius spheres centered on the Voronoi-detected peaks or use the 
percolation algorithm.
In the following, I will describe a 3D technique in which this last reconstruction step
is consistently carried out using the dual structure of a Voronoi partition, i.e. the 
Delaunay tessellation. 

\begin{figure}
\centering
\includegraphics[width=8cm, scale=2., angle=-90]{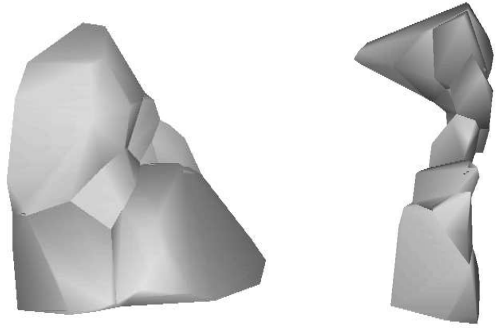}
\caption{ Three-dimensional Voronoi reconstruction of a cluster with 10 galaxies
in a galaxy simulation. The Voronoi cells encompassing each cluster galaxy are shown 
in real space ({\em bottom}) and in redshift space
({\em top}).  Each Voronoi 3D cell surrounding a galaxy is defined as
the intersection of the planes which are perpendicular bisectors of
the lines joining that galaxy to its neighbors. Note how the nearly isotropic
real-space distribution of cluster galaxies degenerates into a
composite Voronoi structure which is elongated along the observer's
line of sight.}
\label{fig:3dv}       %  
\end{figure}

\subsection{The  3D Voronoi and Delaunay Method}

The  Voronoi-Delaunay Method we developed (VDM, \cite{mio})
makes use of the Voronoi partition and Delaunay triangulation 
to identify and reconstruct galaxy clusters in 3D  redshift surveys (see Fig. \ref{fig:3dv}).   
Once the Voronoi/Delaunay calculation for a 3D 
catalog of galaxies has been completed, the algorithm proceeds in three phases.  First, peaks
in density (obtained by sorting the inverse volumes into a monotonic
sequence) are identified and provide candidate locations for cluster
centers.  The Delaunay mesh then allows to identify central members
of each candidate group and estimate physical properties such as the
cluster central density. Finally, these estimates are used to define
redshift space windows within which we find each group's members.  In
this last step it is the predicted structure of the clusters, inferred
from an initial level of grouping, that influences {\em local}
decisions regarding galaxy membership.

\begin{figure}
\centering
\includegraphics[height=9cm]{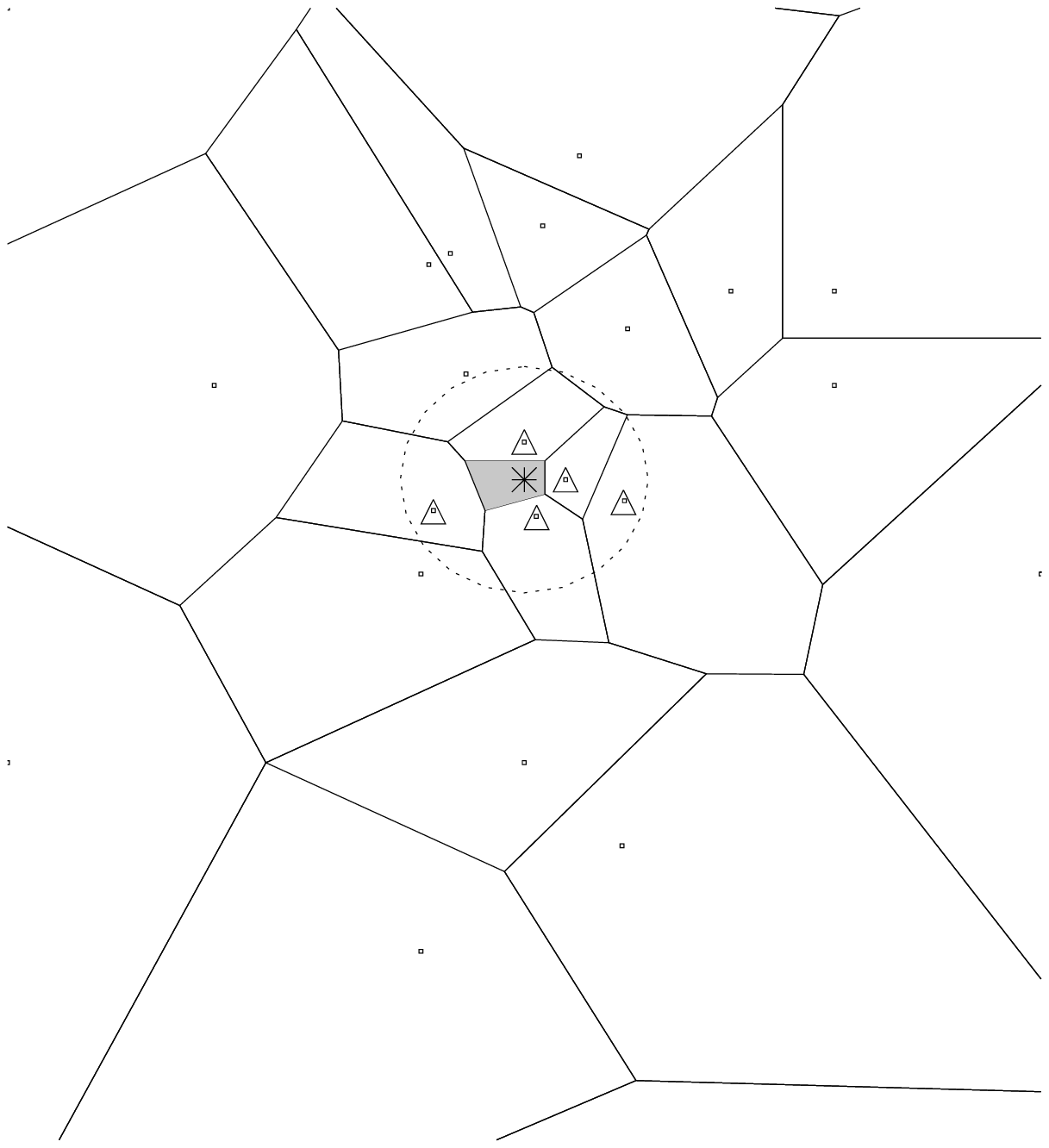}
\includegraphics[height=9cm]{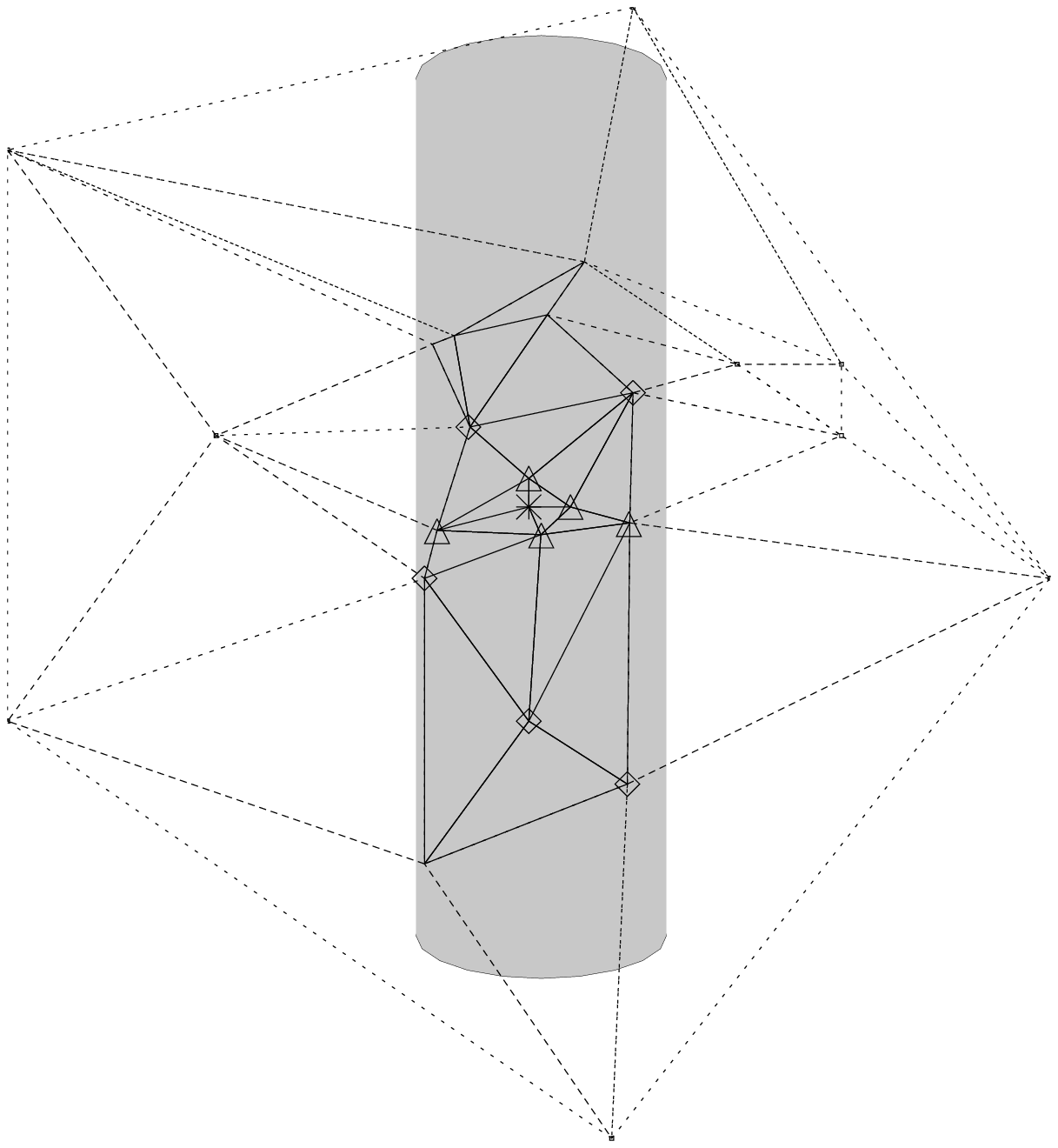}
\end{figure}
\begin{figure}
\caption{{\it Top:} simplified 2-dimensional  graphic representation of phase I of the VDM
algorithm.
Sky angular coordinates are along the x axis, and the survey 
depth is displayed  along the y axis. 
Dots represent individual galaxies, while
the shaded area represents the Voronoi cell surrounding a possible cluster seed
(represented by an asterisk). The set formed by the asterisk and the 
points marked with triangles represents the
{\em first order Delaunay neighbors}.
{\it Bottom} Simplified 2-dimensional graphic representation of phase II (\S 3.2.2). 
Sky angular coordinates are along the x axis, and the survey 
depth is displayed along the y axis.
Segments represent the Delaunay mesh 
connecting the galaxy distribution shown in the top panel. 
{\em First-order Delaunay neighbors} are represented by triangles and {\em second-order
 Delaunay neighbors} by diamonds. Note that not all the galaxies inside the search window 
(shaded area) are used to define the projected  central density parameter
$N_{II}$, but only those designated by symbols. (Taken from \cite{mio}.)}
\label{fig:steps}       %  
\end{figure}

\subsubsection{Phase I: Finding systems of galaxies}

We  begin by assuming that the centers of clusters will lie near peaks
in the galaxy density field.  To
identify these peaks, we sort all the galaxies in the catalog by the
inverse volume of their Voronoi cell; the smallest cells are most
likely to fall at density maxima and are thus potential ``seeds'' for
finding groups or clusters.  We must next determine if a given seed
actually lies at the heart of a system of galaxies.

Each  cluster seed  is   linked to  its nearest  neighbors  by the
Delaunay mesh.  We are interested  only in real, physical groupings of
galaxies; we therefore  must define some {\em ad  hoc} threshold in an
attempt  to  distinguish  galaxies  that  could  be  physically  bound
together from galaxies which are in chance proximity to each other.
We  consider to  be  neighbors those  Delaunay-connected points  whose
distance from  the seed  galaxy is less  than a fixed  limit ${\cal
R}_{min}$.  These  galaxies, and  the original seed  galaxy itself,
will be referred to  hereafter as {\em first-order Delaunay neighbors}
and are used to determine the system's center of mass.

If no  galaxies satisfy this criterion  then the cluster  seed will be
rejected and considered an isolated galaxy.  If, when analyzing a seed
we find that all its {\em first-order Delaunay neighbors} have already
been assigned  to another structure,  we automatically merge  the seed
galaxy  into that system.   A schematic  representation of  this first
step in cluster identification is presented in the top panel of Fig. \ref{fig:steps}.

Because there is a fairly tight correlation between cluster richness
 and the order in which they are reconstructed (as we proceed from the
 highest density cells to the lowest, the richest clusters are
 generally identified first), we need not test every object as a
 potential cluster center.  Instead, in the interest of computational
 speed we exclude the cells around galaxies that have already been
 assigned to systems from being themselves considered as other cluster
 seeds or members.  We thus proceed through the catalog in order of
 increasing Voronoi cell volume, identifying every object in the
 clusters through the three-phase process described above and then
 moving to the next smallest Voronoi cell as yet unclustered, until
 the entire catalog has been either assigned to a cluster or tested as
 a potential cluster seed.

\subsubsection{Phase II: Determining clustering strength} 

Since they are calculated in a parameter-free fashion, both the
Voronoi diagram and Delaunay complex are determined isotropically in
the angular and redshift directions.  However, the peculiar velocities
induced by a cluster's gravitational field cause the distribution of
galaxies to appear elongated in the redshift direction to a degree
determined by its velocity dispersion.  The three-dimensional
information lost in the transformation to redshift space cannot be
recovered uniquely via isotropic, geometric methods; additional
assumptions are required to minimize contamination by spurious
members.  To determine the properties of clusters with any accuracy,
we require methods that include this anisotropy.

We therefore  define a cylindrical window in  redshift space (centered
on the  barycenter determined in Phase  I and circular  in the angular
dimensions) within which we may  find objects which are very likely to
be members  of each  cluster.  This cylinder  will have  radius ${\cal
R}_{II}  \geq {\cal R}_{min}  $ and  length (along  the redshift
direction) ${\cal  L}_{II}$.  We  define those galaxies  which fall
within  this  cylinder  and  are  connected  to  first-order  Delaunay
neighbors  by   the  Delaunay  mesh  as   {\em  second-order  Delaunay
neighbors}; see the bottom panel of Fig. \ref{fig:steps} for a graphical illustration. Note 
that not all the galaxies in the cylinder are included.

We set ${\cal R}_{II}= {\cal R}_{min}=1 ^{-1}$Mpc comoving, which
is the typical central radius theoretically predicted for massive 
and virialized clusters.  Analogous physical considerations guide us to set the
half-length of the cylindrical window, ${\cal L}_{II}$, to be $20
 h^{-1}$Mpc. This value includes the maximum line-of-sight peculiar velocity of galaxies
that are identified as members of systems in N-body simulations 
(as high as $\sim 1500 km/s$).

We then use the sum of the numbers of {\em first- and second-order
Delaunay neighbors} as an indicator of the central richness of the
group, $N_{II}$.  By including only Delaunay
neighbors in $N_{II}$, we are able to minimize contamination by
interlopers, providing a robust estimate even in low-density systems.
This is particularly important because $N_{II}$ controls the
adaptive window for cluster members used in Phase III.

\begin{figure}
\centering
\includegraphics[width=6.6cm, angle=-90]{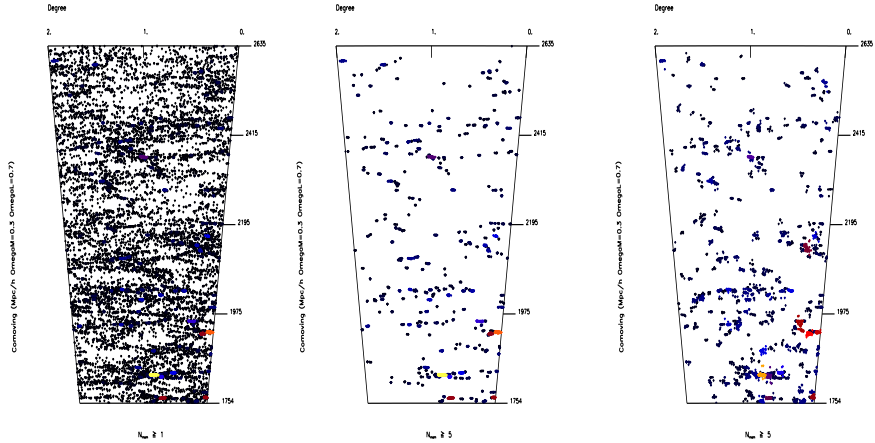}
\caption{{\em Left:}  
N-body simulation of galaxies represented in a  2D, real-space cone diagram. 
The volume  corresponds to an angular area 
of 1 square deg and to the redshift interval z=0.7-1.2 (here expressed 
in Mpc units). {\em Center:} real space, large-scale spatial distribution 
of galaxies belonging to clusters with more  than 5 members as identified 
in the simulation.  {\em Right:} real space, large-scale spatial distribution of 
galaxies belonging to clusters with more 
than 5 members as reconstructed by the VDM algorithm applied in redshift space.
(Taken from \cite{mio}.)}
\label{fig:err1}       %  
\end{figure}

\subsubsection{Phase III: Assigning cluster members} 

Having identified the center and estimated the central richness for each
cluster, we then reconstruct the full set of system members.  We do
this on the basis of physical considerations, not via an empirically
tuned parameter threshold.  In particular, we exploit the known
richness-velocity dispersion correlation to define a search window for
each cluster's members based upon its richness.

The virial theorem predicts that velocity dispersion and central
number density of galaxies are correlated.  \cite{bah}
observationally confirmed the existence of a strong linear correlation
valid from loose groups to clusters.   We rely on this relation to
estimate the strength of the underlying clustering, which we then use
to determine on a group by group basis the window around the system's
center within which to search for Delaunay-connected galaxies.  
Thus, the algorithm relies on the principle that cluster reconstruction
should not proceed by linking a chain of "friends" that could lead to any
given galaxy in the sample, but instead should iterate the search for
cluster members from a potential cluster center position outward in 
an adaptive fashion.

Specifically, for each cluster we define a cylindrical window
(symmetric about the redshift direction)  with radius 
${\cal R}_{\rm II}$ and length $2{\cal L}_{\rm II}$ 
determined by the richness scale factor $N_{II}$ as follows

\begin{eqnarray}
{\cal R}_{\rm III} &=& r(\tilde N_{\rm II})^{1/3}\\
{\cal L}_{\rm III} &=& l(\tilde N_{\rm II})^{1/3} f(z)
\end{eqnarray}

\noindent whereas $r$ and $l$ are two free parameters, $\tilde N_{\rm II}$ is the central 
richness corrected for the redshift dependent mean density $\langle n \rangle(z)$, and $f(z)$ is a function 
introduced to take into account that for a fixed velocity dispersion the length of the fingers 
of god in redshift space is a function of redshift. $\tilde N_{\rm II}$ and $f(z)$ are given 
(in an arbitrary cosmology of matter density $\Omega_m$ and vacuum energy density 
$\Omega_{\Lambda}$) by

\begin{eqnarray}
\tilde N_{\rm II} = \frac{\langle n \rangle(z_{\rm ref})}{\langle n \rangle(z)} N_{\rm II}
\end{eqnarray}
and
\begin{eqnarray}
f(z) = \frac{s(z)}{s(z_{\rm ref})},\ \ \ s(z) = \frac{1+z}{\sqrt{\Omega_m(1+z)^3+\Omega_\Lambda}}
\end{eqnarray}
respectively, where $z_{\rm ref}$ is an arbitrary reference redshift. 

By using the Delaunay mesh to identify the nearby galaxies, we
are able to reconstruct group membership quite rapidly; once the initial Voronoi-Delaunay
calculation is complete (which need only be done once for a catalog),
it takes only 1 minutes on a modern workstation to process $\sim
15000$ galaxies into a catalog of groups and clusters.

\subsubsection{Tests of the VDM algorithm}
What is more important, and what is not usually
done for other methods,  the performances of the algorithm have been 
tested using N-body simulations of 
deep redshift  surveys. This allows a quantitative  characterization  
of  the completeness of the resulting group catalog. 

In particular  Gerke et al. (\cite{ger})  found that the group-finder 
can successfully identify $\sim78\%$ of real groups and that $\sim79\%$ of the
galaxies that are true members of groups can be identified as 
such.  Conversely, $\sim 55\%$ of the groups found can be definitively 
identified with real groups and $\sim46\%$ of
the galaxies placed into groups are interloper field galaxies.

\begin{figure}
\centering
\includegraphics[height=13cm]{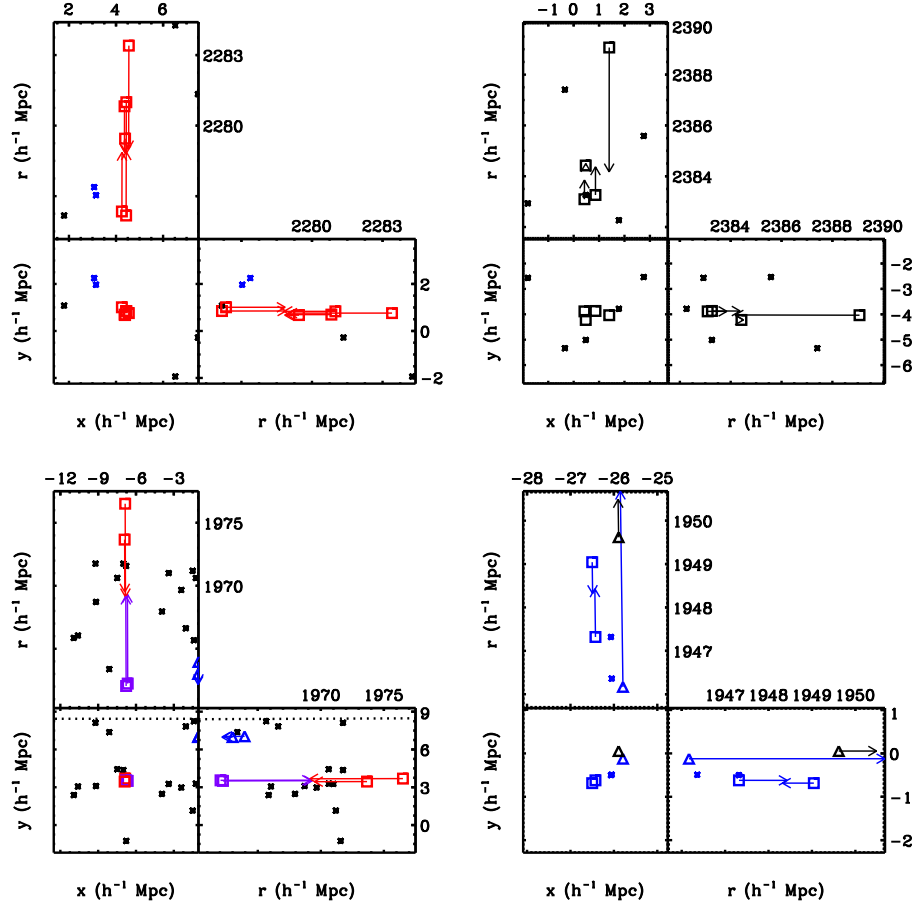}
\caption{Examples of group-finding success and failure \cite{ger}. 
In each panel, squares indicate galaxies 
in a simulated group being plotted, triangles indicate galaxies in nearby real groups, 
and crosses indicate nearby field galaxies. Galaxies are plotted both as seen on the sky 
and in two line-of-sight projections in redshift space. Arrows point to the galaxies  real 
space positions, to show the effects of redshift-space distortions (to reduce visual clutter,
 no arrows are plotted for field galaxies). Dotted lines indicate field edges. 
Each reconstructed group is indicated by a different shade, with reconstructed field 
galaxies shown in black. The top left panel shows a perfect reconstruction with a nearby
 false detection; the top right panel shows a completely undetected group (all black squares), 
the bottom left panel shows a fragmented real group, and the bottom right panel shows an 
example of over-merging. (Taken from \cite{ger}.)}
\label{fig:err2}       %  
\end{figure}

To give a visual  sense of the errors encountered in 
reconstructing individual groups, we show several examples of
group-finding success and failure in figures \ref{fig:err1} and \ref{fig:err2}.

\begin{figure}
\centering
\includegraphics[height=13cm]{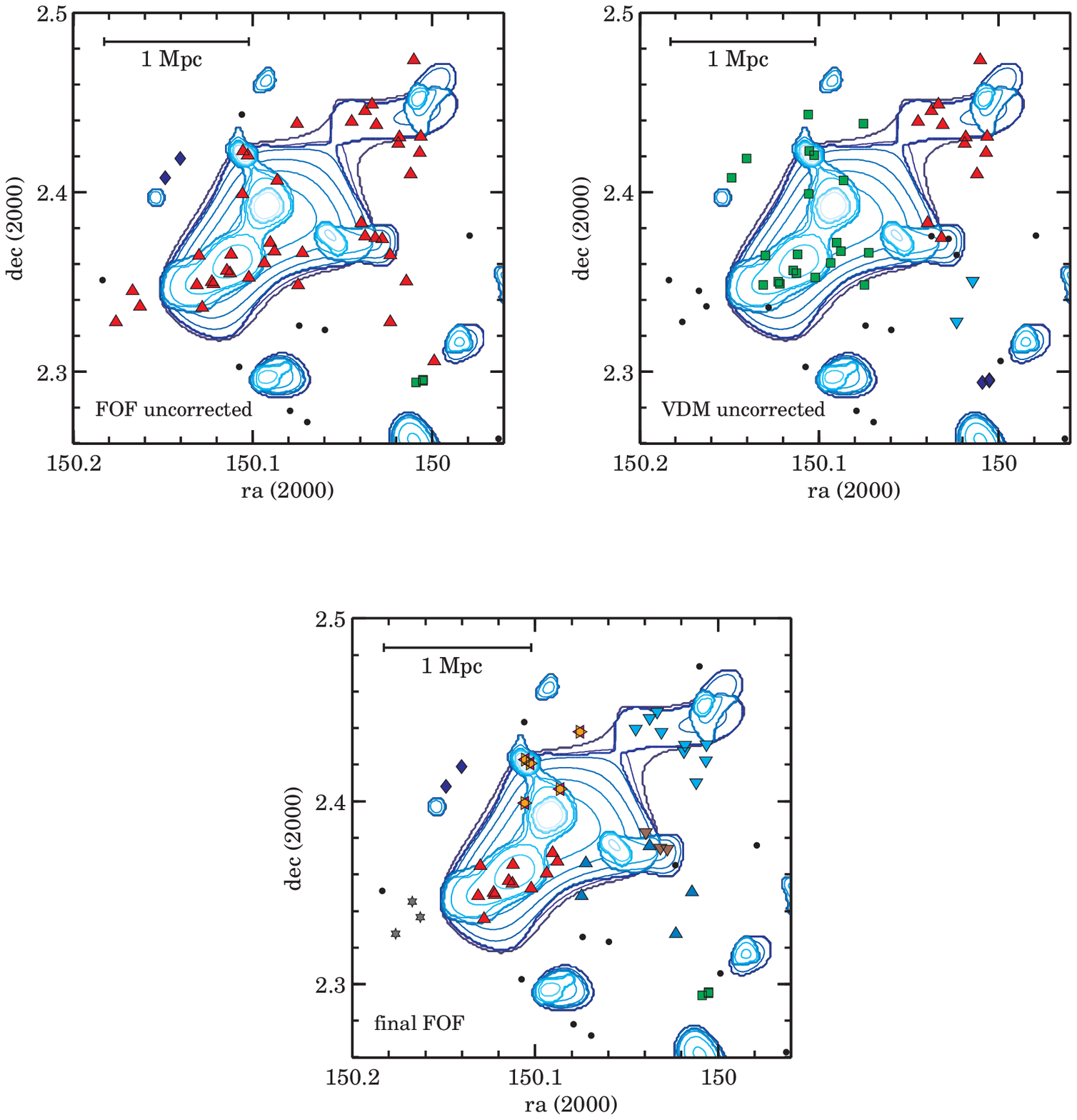}
\caption{The upper left panel shows the structure as reconstructed by the FOF method, 
the upper right the VDM recovered systems.
Because of the failure of the FOF, the group recovered by this method  has been a-posteriori 
manually split up into several groups (lower panel). 
Black points denote field galaxies, and the other symbols (squares, triangles, etc.) are 
group galaxies, whereas each group has its own symbol and color. The blue contour exhibit 
the X-ray emission of the super-group as observed with XMM-Newton.
Taken from \cite{knoebel}. \label{fig:kn}}
\end{figure}

A compared analysis of the VDM and FOF performances has also been carried out \cite{knoebel}
and the main result is graphically summarized in figure \ref{fig:kn} where the performances
of the two methods in recovering a particular structure are contrasted.
This structure
is probably an example of a ``super-group'', where several smaller groups are just 
about to merge \cite{fino}. 

The upper left panel shows the group assignment of the FOF method, and the upper right panel 
the group assignment of the VDM. Each group is denoted by a symbol (e.g. square, triangle) 
of a particular color, and field galaxies by black points. This example of the super-group 
gives us some interesting insights concerning the group-finding procedure.

This extended structure exhibits the main potential problems of both the FOF and VDM algorithms. 
The FOF algorithm connected practically all the galaxies in this super-group, 
without distinguishing between different sub-groups. As discussed, this behavior is well 
known for FOF, and it happens in particularly dense regions such as this. 
The problem is that any single galaxy between two of these sub-clusters will act as a 
bridge for the FOF algorithm to connect the two clusters. The VDM is more successful 
in distinguishing different sub-structures, but nevertheless fails to do a perfect job. 
A casual glance suggests that the ``green square'' VDM cluster in fact consists of two 
independent sub-groups (consistent with the X-ray contours). 
Furthermore, the ``red triangle'' VDM group exhibits two outliers to the South which almost 
certainly do not belong to this group. The occurrence of such outliers is not uncommon in VDM 
groups. It is related to the fact that in the VDM group-finder every second order 
Delaunay-neighbor in the second cylinder is accepted as group member and that the second 
cylinder is usually much bigger than the third cylinder \cite{knoebel}.

Since, as we have seen, any group-finding algorithm is prone to many different types of 
systematics,  it is crucial to define carefully the tolerance for various errors and craft 
a specific  definition of group-finding success.  
We focused our efforts on reproducing as accurately as possible 
only those specific group properties which are more relevant 
for cosmological purposes. In particular Marinoni et al. and Gerke et al. 
showed  that  it is possible to quantify  
the completeness of the resulting group catalog. 
directly in the physical space of velocity dispersions. In other words,
it is possible to define a velocity dispersion threshold $\sigma$ above which 
a critical cosmological statistics such as 
the spatial abundance of systems  at a given redshift ($N(>\sigma,z)$)
is recovered in a complete way with a sufficient degree of accuracy.

A common feature of most of the algorithms which make use of the Voronoi
partition to identify cluster candidates in 2D, is that cluster members identification 
is usually carried out using additional and independent methods. For example, the 
most popular technique consists in selecting as clusters members those galaxies 
which lie within fixed-radius spheres centered on the Voronoi-detected peaks or use the 
percolation algorithm.
On the contrary in the 3D VDM technique this last reconstruction step
is consistently carried out using the dual structure of a Voronoi partition, i.e. the 
Delaunay tessellation. This offers  the possibility of adaptively
scaling cluster selection window only on the basis of the geometrical characteristics
of the Voronoi and Delaunay meshes, i.e. on the basis 
of completely non-parametric structures.

I will emphasize three other aspects of the VDM algorithm.
{\it a)} The method is based on the specific idea that cluster
reconstruction does not proceed by linking potential friends to any
given galaxy in the sample (such as in the standard 3D percolation or 
hierarchical methods), but by iterating the search of cluster
members from a potential cluster center position outward (which is the natural 
identification process in visual or 2D cluster reconstruction);\\
{\it b)} there is no need to introduce an arbitrarily chosen global  density threshold to
judge when a given system
is formed. Instead the cluster searching window is  locally scaled
on a cluster by cluster basis using physical arguments (in the specific case
the strength of the first Delaunay connected units);

\section{Conclusions}

The VDM algorithm, working with controlled reliability and completeness over a wide range 
of redshifts and a large degree of density enhancements, has been used to identify groups 
and clusters in all the existing deep redshift surveys of the universe:  the DEEP2 \cite{ger},
the VVDS \cite{cucciati}, and the zCOSMOS \cite{knoebel} surveys.

Ongoing analysis, however, shows that, despite remarkable progresses, we have yet 
to design the "perfect" group-finder. While recovering 
rich systems is quite straightforward (especially in 
redshift surveys), reconstructing with high efficiency structures with fewer 
members (i.e. poor groups or poorly sampled clusters) is technically challenging. 
In particular, the level of 
completeness and purity with which groups are currently 
identified ($\sim 80\%$) is still far from optimal. 

A temporary way out consists in  compensating for weaknesses and failures characterizing 
each single method by cross-identifying and matching structures in catalogues 
produced by different algorithms \cite{knoebel}. 
Anyway, developing more sophisticated cluster finding algorithms is crucial 
if we want to reconstruct galaxy environment with high precision 
and use the next generation of deep 3D data to answer key cosmological questions.

It is also critical to stress that the optical identification of a strong 
spatial concentration of galaxies is not a sufficient criterium for the 
identification of a  virialized cluster of galaxies. It is by no means clear whether,
in this way,  one has found a gravitationally bound and relaxed system of galaxies and 
the corresponding dark matter halo. Even centrally condensed structures, as far as the optical 
distribution of galaxies is concerned, may show relevant substructures 
in the diffuse gas distribution when their X-ray image is analyzed. For example, the 
Coma cluster, long considered to be a regular cluster, is not completely in an equilibrium
state, but it is dynamically evolving presumably by the accretion of an adjacent galaxy group.  

With this caveat in mind, it is encouraging to turn the head backward and appraise
the long way walked from the very early days dominated by a sort of skeptical attitude
towards this cosmographical endeavor
up to an epoch, the present, in which  we can reliably use catalogs of optical 
groups an clusters to gain access to a wealth of crucial  cosmological  information.

\acknowledgement

I'd like to thank C. Adami, A. Biviano, A. Iovino, A. Mazure and R. Van de Weygaert for their careful 
reading of the manuscript and helpful comments. 

%%%%%%%%%%%%%%%%%%%%%%%% referenc.tex %%%%%%%%%%%%%%%%%%%%%%%%%%%%%%
% sample references
% "physics"
%
% Use this file as a template for your own input.
%
%%%%%%%%%%%%%%%%%%%%%%%% Springer-Verlag %%%%%%%%%%%%%%%%%%%%%%%%%%

%
% BibTeX users please use
% \bibliographystyle{}
% \bibliography{}
%
% Non-BibTeX users please use

\printindex
\end{document}